\documentclass[a4]{article}
\usepackage[english]{babel}
\usepackage[utf8]{inputenc}
\usepackage[mathletters]{ucs}
\usepackage{amsmath}
\usepackage{graphicx}
\usepackage{subcaption}
\usepackage{epsfig}
\usepackage[colorinlistoftodos]{todonotes}
\usepackage[hidelinks]{hyperref}
\usepackage[margin=1in]{geometry}
\usepackage{multicol}
\usepackage{caption}[=v1]
\usepackage{lmodern}
\usepackage{siunitx}
\usepackage[super,comma,sort&compress]{natbib}

\usepackage{multirow}
\usepackage{booktabs}
\usepackage{authblk}
\usepackage{xr}
\usepackage{setspace}

\title{Kinetic control of ferroelectricity in ultrathin epitaxial Barium Titanate capacitors}

\date{}
\author[1]{Harish Kumarasubramanian}
\affil[1]{Mork Family Department of Chemical Engineering and Materials Science, University of Southern California, CA 90089, USA }
\author[2]{Prasanna Venkat Ravindran}
\affil[2]{School of Electrical and Computer Engineering, Georgia Institute of Technology, Atlanta, Georgia 30332, USA}
\author[1,3]{Ting-Ran Liu}
\author[2]{Taeyoung Song}
\author[1,3]{Mythili Surendran}
\affil[3]{Core Center for Excellence in Nano Imaging, University of Southern California, 925 Bloom Walk, Los Angeles,
CA 90089, USA }
\author[1]{Huandong Chen}
\author[4]{Pratyush Buragohain}
\author[4]{I-Cheng Tung}
\author[4]{Arnab Sen Gupta}
\author[4]{Rachel Steinhardt}
\author[4]{Ian A. Young}

\affil[4]{Technology Research, Intel Corporation, Hillsboro, OR, USA}
\author[1,3]{Yu-Tsun Shao}
\author[2]{Asif Islam Khan}
\author[1,3,5]{Jayakanth Ravichandran}

\affil[5]{Ming Hsieh Department of Electrical Engineering, University of Southern California, Los Angeles,
California 90089, USA}

\graphicspath{Images_new }

\begin{document}
\maketitle

\begin{abstract}
Ferroelectricity is characterized by the presence of spontaneous and switchable macroscopic polarization. Scaling limits of ferroelectricity have been of both fundamental and technological importance, but the probes of ferroelectricity have often been indirect due to confounding factors such as leakage in the direct electrical measurements. Recent interest in low-voltage switching electronic devices squarely puts the focus on ultrathin limits of ferroelectricity in an electronic device form, specifically on the robustness of ferroelectric characteristics such as retention and endurance for practical applications. Here, we illustrate how manipulating the kinetic energy of the plasma plume during pulsed laser deposition can yield ultrathin ferroelectric capacitor heterostructures with high bulk and interface quality, significantly low leakage currents and a broad "growth window". These heterostructures venture into previously unexplored aspects of ferroelectric properties, showcasing ultralow switching voltages ($<$0.3 V), long retention times ($>$10$^{4}$s), and high endurance ($>$10$^{11}$cycles) in 20 nm films of the prototypical perovskite ferroelectric, BaTiO$_{3}$. Our work demonstrates that materials engineering can push the envelope of performance for ferroelectric materials and devices at the ultrathin limit and opens a direct, reliable and scalable pathway to practical applications of ferroelectrics in ultralow voltage switches for logic and memory technologies.

\end{abstract}

\normalsize 
\section*{Introduction}
\par

The scaling limits of ferroelectricity have been an open question with both scientific and technological importance. Over the past five decades, the advances in modeling and experimental methods have resulted in a greater understanding of ferroelectricity in ultrathin films, especially in prototypical perovskite ferroelectrics\cite{dawber2005physics,stengel2006origin,fernandez2022thin,sai2005ferroelectricity,junquera2003critical}. However, significant challenges remain in achieving robust ferroelectric properties such as long retention times and high endurance, even in epitaxial, single crystalline perovskite ferroelectric films at the ultrathin limit\cite{kim2005polarization,ma2002nonvolatile}. Here, both intrinsic and extrinsic factors such as depolarization effects and defect induced leakage limit our ability to achieve and sustain ferroelectricity. The relationship between electrically active defects and growth processes remains poorly understood hampering the means to mitigate these extrinsic factors. While fluorite-structured binary oxides and wurtzite-structured nitrides have emerged as interesting ferroelectric materials, their large coercive fields \cite{kim2023wurtzite,mikolajick2021next} limit their low-voltage operation, an important requirement for advanced, ultra-low power, high-performance electronics. This fundamentally limits our ability to take advantage of the low-power capabilities of ferroelectric transistors over existing transistors with dielectric gates \cite{hoffmann2021progress}. Hence, there is an urgent need to advance the understanding of growth processes towards mitigating extrinsic effects that limit ferroelectricity in perovskite ferroelectric thin films. Recently, dramatic improvements in low voltage switching of capacitors with a perovskite ferroelectric, BaTiO$_{3}$ were achieved by carefully balancing the processing and the strain state\cite{jiang2022enabling}, but the reliability of these films at ultrathin limits remains an outstanding challenge. Further advances are critical to realize low voltage\cite{manipatruni2018beyond}, electronic switches for logic and memory applications.

In the last two decades, reports have shown evidence of ferroelectricity in perovskite films only a few unit cells thick in the ideal Metal-Insulator-Metal (MIM) geometry\cite{gao2017possible,lee2019first,nagarajan2006scaling,shaw2000properties,jo2005thickness,junquera2003critical}. Unfortunately, many of the studies on ferroelectricity have relied on indirect methods such as X-ray or optical probes that interpret polar distortions of the lattice as evidence for ferroelectricity. Although the polar nature of ferroelectric materials survive down to the unit cell level, several factors such as leakage currents and depolarization fields quench the macroscopic polarization. Furthermore, when ferroelectricity is inferred in ultrathin films by local probing techniques such as Piezoresponse Force Microscopy (PFM), the “polar” response could arise from extraneous electrochemical, electromechanical, and/or electrostatic effects and not exclusively limited to ferroelectric origins \cite{seol2017non,vasudevan2017ferroelectric}. Additionally, these techniques are non-quantitative when it comes to measuring device parameters such as polarization, and coercive field. Even if these films are “switchable” in regions spanning few hundreds of nanometers, the absence of a large area, homogeneous and macroscopic polarization response limits the reliability of these approaches for use in device applications. Considering the concerted effort to integrate these materials into commercial semiconductor devices, ferroelectricity in ultrathin perovskites must be achieved in the truest sense – a static, spontaneous, switchable dipole accessible over macroscopic dimensions, to enable reliable devices based on ferroelectrics. In this article, we show that careful control of the kinetic energy of the deposited species in Pulsed Laser Deposition (PLD) paves the way to overcome nearly all the extrinsic limitations in epitaxial ferroelectric heterostructure capacitors with Barium Titanate (BTO) and, hence a direct pathway for ultralow voltage switches and electronic devices based on ferroelectrics. More importantly, this insight broadens the growth window, where robust ferroelectric heterostructure capacitors can be achieved, potentially opening up PLD for high-volume manufacturing of ferroelectric materials with complex chemistries.

\par

\begin{figure*}
 \centering
\includegraphics[width=0.85\linewidth]{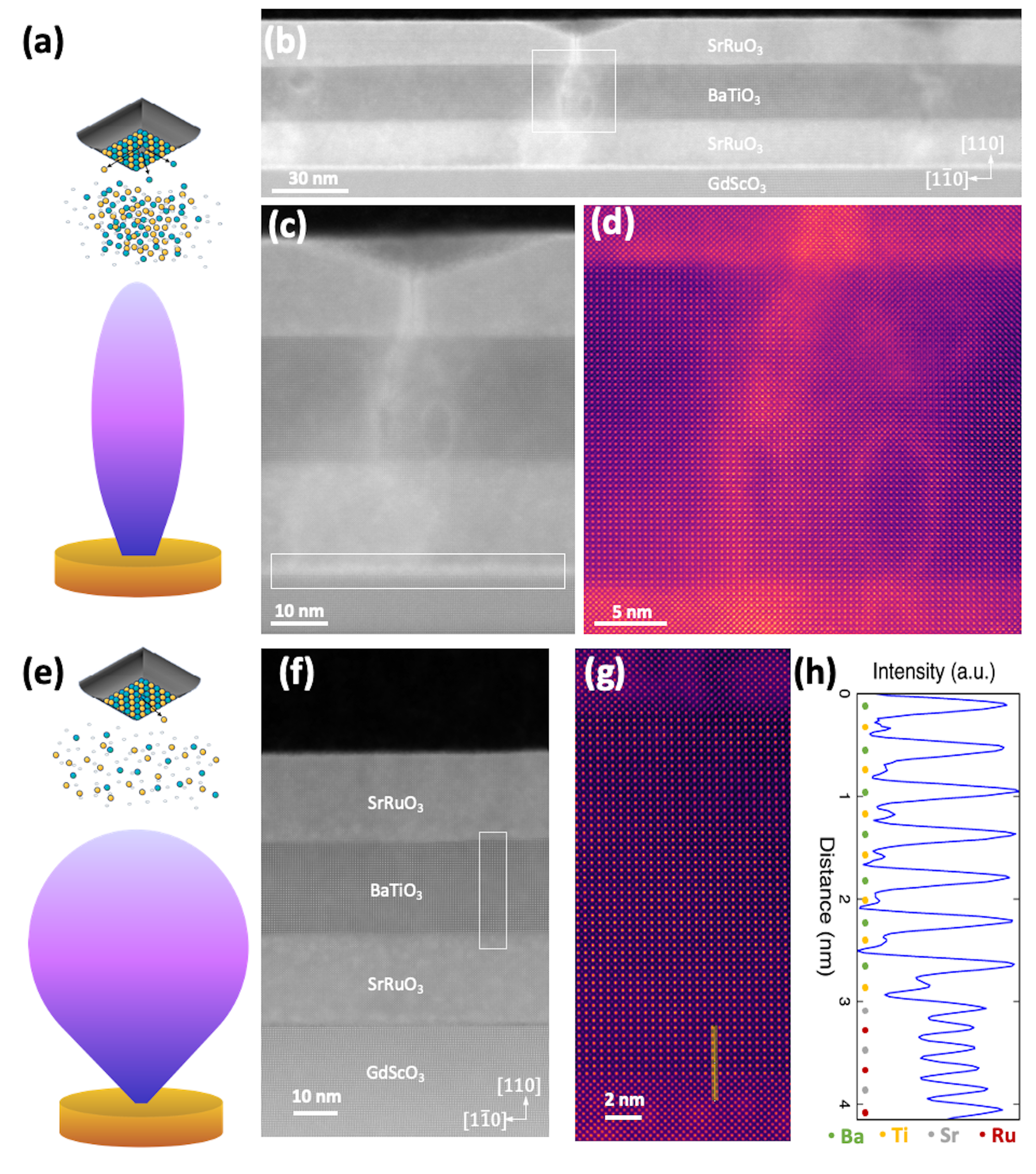}\captionof{figure}{The effect of (a) a diffuse laser spot with a large area and (e) a focused laser spot with a smaller area on the plume dynamics and the kinetic energy of the species provided that all other growth parameters, including laser fluence, remain the same. The bombardment effects on the film can be greater with the larger spot, as illustrated in (a). (b)-(d) shows the HAADF-STEM image of a 20 nm SrRuO$_3$/20 nm BaTiO$_3$/20 nm SrRuO$_3$ heterostructure on GdScO$_3$ substrate where in the spotsize and the plume dynamics were akin to the one shown in (a). Utilizing a low angle annular dark field (LAADF) with the collection angle from 33 to 199 mrad, the defects were examined throughout the 20 nm SrRuO$_3$ /20 nm BaTiO$_3$/20 nm SrRuO$_3$ heterostructure. (b) Cross-sectional view on large area of the heterostructure reveals multiple defective regions. Notably in (c), the bottom SrRuO$_3$/GdScO$_3$ features a thin layer of distorted atomic columns, depicted in the white box. (d) High-resolution image of the BaTiO$_3$/SrRuO$_3$ interface from the labeled region in (b), indicated by the white box. Furthermore, the BaTiO$_3$ layer has defective atomic columns. (f)-(h) shows the HAADF-STEM image of a 20 nm SrRuO$_3$ /20 nm BaTiO$_3$/20 nm SrRuO$_3$ heterostructure on GdScO$_3$ substrate where in the spotsize and the plume dynamics were akin to the one shown in (e). (f) Cross-sectional view of a large area shows a sharp interface on both sides of the BaTiO$_3$/SrRuO$_3$ layers. No obvious crystallographic defects were found over a 10 um field of view. (g) High resolution image of the BaTiO$_3$/SrRuO$_3$ interface from labeled region in (f), white box. (h) Intensity profile of the HAADF image from yellow line in (g) showing the atomically sharp interface.}
 \label{fig:SS,STEMandXRD}
\end{figure*}
\par

\subsection*{Processing Challenges for Ferroelectric Thin Films}

Polarization in a ferroelectric is affected by many intrinsic and extrinsic factors. Intrinsic factors limit the ferroelectric properties even in idealized "defect-free" conditions given the choice of the ferroelectric and the electrode material. For instance, the low potential barrier against back switching for perovskites\cite{gong2016fe}, the lower conductivity/higher Thomas-Fermi screening length for epitaxial perovskite electrodes\cite{black1999electric} and the resultant depolarization fields are intrinsic factors, as they are intrinsic to the choice of materials. There is a fundamental correlation of increase in this depolarization field with decreasing thickness that will ultimately limit the stability of any ferroelectric phase. On the other hand, one may overcome extrinsic factors such as control of defects that lead to leakage or depolarization effects. A variety of vapor phase growth techniques, such as molecular beam epitaxy (MBE), pulsed laser deposition (PLD), and sputtering, have been used to grow thin films of model perovskite ferroelectric materials such as BaTiO$_{3}$, PbTiO$_{3}$, and BiFeO$_{3}$. Despite sustained advances in growing stoichiometric thick films ($>$100 nm) with excellent ferroelectric properties, the ability to achieve thin films ($<$50 nm) with good ferroelectric properties remains an open challenge due to narrow and often, inconsistent and irreproducible growth windows for the deposition of these materials\cite{martin2024lifting}. Ferroelectricity in thinner perovskite films ($<<$ 50 nm) depends not only on the ability to control the stoichiometry but also on the nature and concentration of defects in these films. In PLD, the control of deleterious defects is closely related to the kinetic energy of the plasma plume and has usually been approached by optimizing various growth parameters such as laser fluence\cite{xu2013impact} and background gas pressure\cite{scullin2010pulsed}. Here, we show a series of experiments that aim at delineating the complex interplay between the different growth parameters and their effects on the functional properties of ultrathin perovskite ferroelectrics.

In addition to the laser fluence and the background gas pressure, an underutilized growth knob is the spot size of the laser on the target. Interestingly, there have been very few, but consistent reports on controlling the growth kinetics and the plume energetics by only varying the laser spot size, but maintaining a constant magnitude of laser fluence\cite{thompson2016enhanced,saremi2016enhanced}. Figures \ref{fig:SS,STEMandXRD} (a) and (e) describe the effects of larger and smaller laser spot sizes on the plume dynamics and ultimately the kinetic energy (KE) of the plasma plume. The smaller, more focused spot results in a more divergent plasma plume with a larger scattering cross section and lower flux per unit area. For the same background gas pressure and laser fluence, the plume is scattered much more, resulting in a perceptible decrease in the KE of the particles in the plume reaching the substrate. As the plasma plume in the case of a larger spotsize is more directional, the particles are scattered less by the surrounding gas molecules. This leads to a higher KE of the ions in the plasma and results in greater resputtering/bombardment effects on the film surface, as illustrated in Figures \ref{fig:SS,STEMandXRD} (a) and (e). Figures (b)-(d) and (f)-(h) show the Scanning Transmission Electron Microscopy (STEM) images for representative heterostructures of 20 nm SrRuO$_3$/20 nm BaTiO$_3$/20 nm SrRuO$_3$/GdScO$_3$ grown using the large and small spotsizes respectively. Circling back to the relationship between the KE of the species in the plasma plume, the heterostructure deposited in the higher KE condition has more extended defects such as dislocations, while the heterostructure grown in the lower KE condition has a defect-free and atomically sharp interface. The interconnection between the functional properties and the spotsizes is discussed in detail in the later sections but essentially, it is this additional control lever that we exploit in this work to improve the bulk and interfacial structural quality of the ferroelectric layer to achieve robust retention and endurance characteristics in ultrathin perovskite ferroelectrics.

\par

\section*{Results and Discussion}

\begin{figure}
 \centering\includegraphics[width=\linewidth]{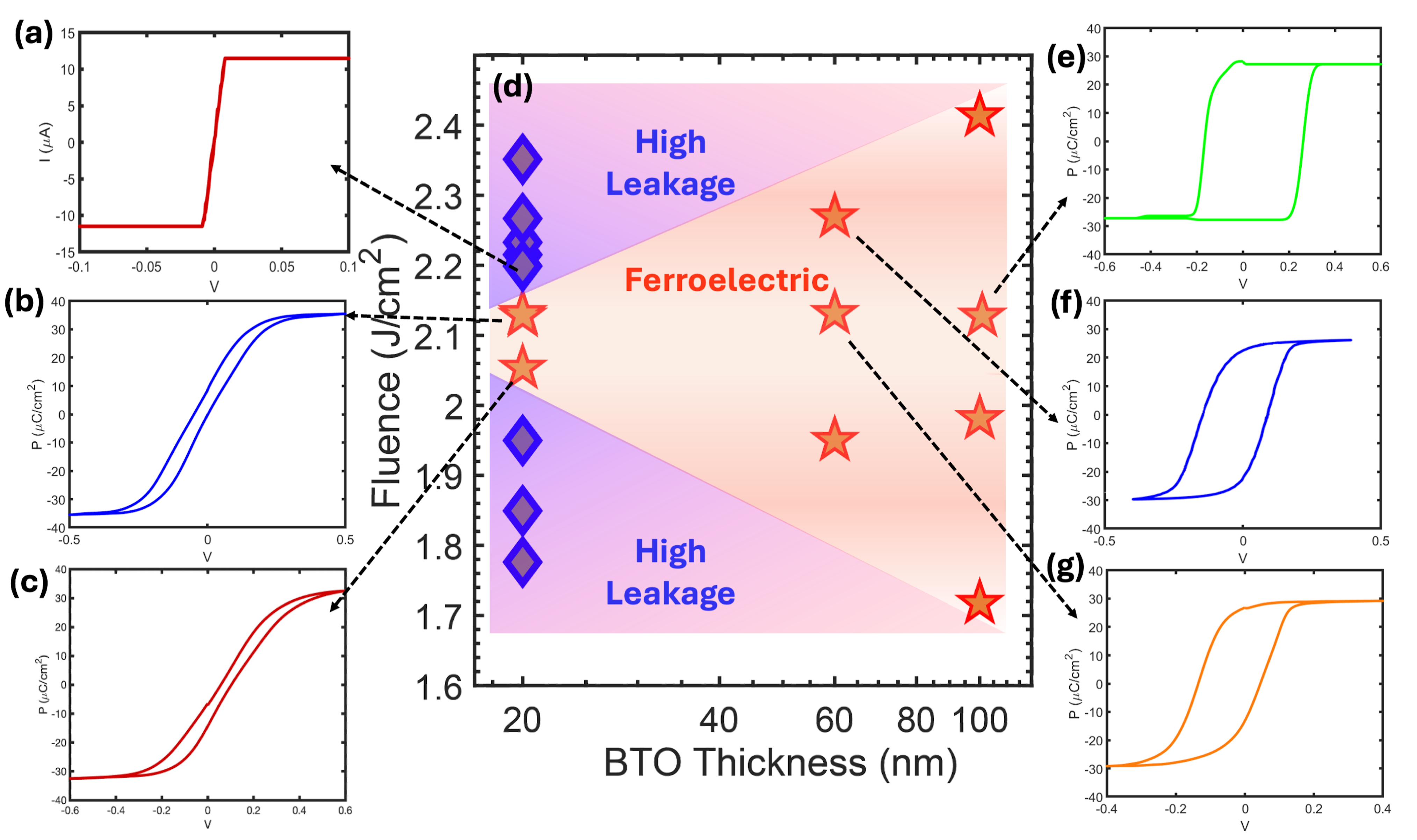}
 \captionof{figure}{(a) Large leakage currents in 20 nm BTO films. (b) and (c) show P-V loops for ferroelectric 20 nm BTO films. (d) Phase space of fluence and thickness that renders ferroelectric films. (e) and (f)-(g) show P-V loops for ferroelectric 100 nm and 60 nm BTO films respectively. All BTO layers were grown at 0.11 mbar oxygen partial pressure.}
  \label{fig:fluence}
\end{figure}

Epitaxial BTO films of various thicknesses (20-100 nm) were grown on single crystal GdScO$_3$ (110) (GSO) substrates. All BTO films were sandwiched between two epitaxial layers of SrRuO$_3$ (SRO) metal electrodes with a thickness of approximately 20 nm. The GSO substrates were chosen because of their close lattice match with BTO\cite{biegalski2005thermal} and their ability to exert a small bi-axial in-plane compressive strain (0.78\%), which favors only a slight increase in the bulk Curie temperature, retaining the potential for small coercive fields and low switching voltages. The structural and ferroelectric characteristics of the heterostructures as a function of different BTO thicknesses are shown in Figures S4 and S5 and are discussed in the supporting information.

\subsection*{Growth Window for Ferroelectricity}

The growth conditions for the SRO/BTO/SRO/GSO heterostructures were initially optimized on 100 nm BTO films. Films grown at 0.11 mbar showed 2D growth mode (figure S2), and correspondingly most of the films grown at this pressure were ferroelectric at a range of laser fluences (figure S3). Hence, this pressure was chosen to deposit thinner BTO films. Figure \ref{fig:fluence}(d) shows the effect of laser fluence on the ferroelectric properties of BTO as a function of thickness. It can be seen that the 100 nm films are ferroelectric for a wide range of fluence values. As the BTO film thickness decreases, the window of fluence for low leakage ferroelectric films becomes narrow. Presumably, there is a smaller margin for the formation of defects that lead to enhanced leakage and/or suppression of ferroelectricity. These extrinsic sources of leakage coupled with exceptionally high depolarization could explain why it has been extremely challenging to achieve reliable ferroelectricity in ultrathin perovskite films.

\par
The effects of varying the laser fluence\cite{kan2011controlled,xu2013impact} and background gas pressure\cite{chen2013strong} on the structure and ferroelectric properties of BTO have been extensively studied and optimized for thick BTO films. As shown in Figure \ref{fig:fluence} in our case, the narrowing growth window due to the combined effects of defect formation and depolarization, achieving a reliable ferroelectric response in ultrathin films remains a challenge. Here, we hypothesize that one can balance the defect formation and the kinetic energy of adatoms to promote crystallinity by controlling the kinetic energy (KE) of the species in the plasma plume. Past studies using a Langmuir probe to study the KE of the plasma plume\cite{harris2020geometrical,lee2016growth} have established the laser spotsize as a parameter to control the KE of the plasma. Further, the control of KE of plasma has been used to obtain electrical properties close to single crystals in both SrTiO$_3$ (STO) and SrRuO$_3$ thin films, suggesting the potential for control over both extended and point defects.\cite{lee2016growth,thompson2016enhanced}, For example, STO films grown in very low partial pressure of oxygen, where point and extended defects are widely theorized to occur, were insulating and optically transparent.\cite{lee2016growth} On the other end of the spectrum, superior conductivity, on par with single crystals have been achieved in SRO (the electrode in this study) by similar modifications to the plume and growth kinetics through the spotsize\cite{thompson2016enhanced}. 

\par

\begin{figure}
 \centering\includegraphics[width=\linewidth]{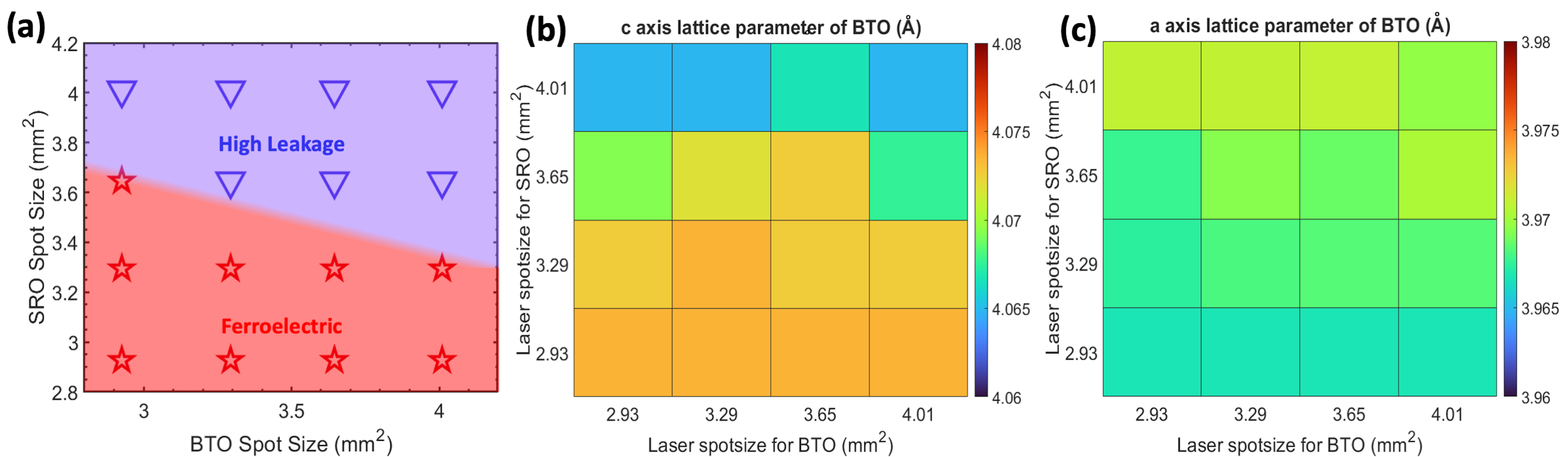}
   \captionof{figure}{(a) Phase space of SRO and BTO spot sizes rendering ferroelectricity in 20 nm SRO/20 nm BTO/20 nm SRO/GSO heterostructures. All BTO films were grown at a laser fluence of 2.13 J/cm$^2$ and a pressure of 0.11 mbar. (b) c axis, (c) a axis lattice parameters for the 20 nm BTO films grown at different spotsizes.}
 \label{fig:BTOc_a_PUND}
\end{figure}

We studied the impact of plume kinetic energy (KE) on the ferroelectric properties of thin BTO, specifically 20 nm thick BTO films sandwiched between 20 nm thick SRO electrodes. During the experiments, to balance the need to achieve stoichiometric materials and the defect formation, the laser spot sizes for both SRO and BTO were varied, while keeping all other growth parameters—such as laser fluence, background gas pressure, and the distance between target and substrate — fixed based on previously optimized values for the respective layers. For both these materials, the deposition rate decreases as the laser spotsize is decreased (as expected from literature reports - Figure S6). The resulting effect on the dielectric properties of the films is depicted in figure \ref{fig:BTOc_a_PUND}(a). Figure \ref{fig:BTOc_a_PUND}(a) was constructed by combining the leakage current and the PUND results shown in Figure S7. Films that exhibit a nonzero switching charge in the PUND analysis (Figure S7(c)) also correspondingly have lower leakage currents. These films are classified as "ferroelectric" in Figure \ref{fig:BTOc_a_PUND}(a). For those samples with high leakage, the P-V loops had to be adjusted using standard Dynamic Leakage Current Correction (DLCC) techniques. In contrast, a distinct loop opening and a non-zero switching charge were observed in both the P-V and PUND analyzes (Figures S7-S12) for the films identified as ferroelectric in figure \ref{fig:BTOc_a_PUND}(a).

\par
There is a clear demarcation in the electrical properties of the films as a function of the plume kinetic energy. It is evident that reducing the SRO spot size below a certain threshold (from 3.65 mm$^2$ to 3.29 mm$^2$ in this study) significantly diminishes the leakage currents by several orders of magnitude. This reduction is critical for stabilizing the spontaneous polarization within these films, thereby ensuring their strong ferroelectric properties. While the SRO spot size emerges as the primary determinant of the ferroelectric characteristics of the heterostructures, the laser spot size used for BTO deposition also has a slight, yet noticeable impact. Presumably, beyond a certain spot size threshold, the increased kinetic energy of the plume and its associated bombardment/sputtering effects are sufficient to cause
elevated leakage currents that overwhelm the ferroelectric properties. We attribute this to the formation of extended defects at higher KE as shown in representative images in Figure 1. The non-trivial and differentiable role of the kinetic energy in controlling the bulk and the interfacial characteristics could be deduced from the slope of the boundary that separates the ferroelectric films from those with high leakage in figure \ref{fig:BTOc_a_PUND}(a). Thus, we have shown that tailoring the characteristics of the metal/ferroelectric interfaces (via SRO deposition) plays a more significant role than bulk modifications (through BTO deposition) in defining the ferroelectric properties of the heterostructure as a whole.

\par
Figure \ref{fig:BTOc_a_PUND}(b) and (c) indicate the structural differences between the BTO films grown in these different kinetic regimes. The in-plane lattice parameters were extracted from a combination of the out-of-plane lattice parameters and by measuring an off-axis reflection of BTO ( 103 reflection). The 002 and 103 reflections are shown in figures S14 and S15 in the supporting information. The range in the $c$-axis lattice parameters between the largest and smallest values for the BTO layers was 0.85 pm with a corresponding variation in $2\theta$ of $0.1^{\circ}$ and the variation in the in-plane lattice parameters ($a $-axis) for the different heterostructures is less than 0.4 pm. Reciprocal Space Maps (RSM) for the two extremes in the spotsize phasespace are shown in figure S16. In both these heterostructures of disparate properties, the SRO and BTO are coherently strained to that of the GSO substrate indicating that the structural variations caused by these different spotsizes are very subtle, atleast in the bulk. Furthermore, the growth modes and morphology for all 16 films in the phase space while varying the spotsizes are all layer-by-layer. A change in the fluence and background gas pressure would result in a discernible change in both the growth mode (observed \textit{in situ} using reflection high energy electron diffraction (RHEED) and the lattice parameter through XRD. On the other hand, electron microscopy studies (figure \ref{fig:SS,STEMandXRD}) show clear and noticeable differences between the films grown in these disparate KE conditions. Reexamining the STEM images in Figure \ref{fig:SS,STEMandXRD}, (b)-(d) represent films grown with the highest SRO spot size - 4.01 mm$^2$ (BTO spot size of 2.92 mm$^2$) in this study. Here, we observe significant distortions in the atomic columns that could possibly be attributed to dislocation or other extended defects. They seem to cause substantial leakage currents in the system and suppress the ferroelectricity. The heterostructures in which the SRO layers were grown under a low KE regime(SRO spot size of 3.29 mm$^2$),were defect-free with sharp interfaces exhibiting robust ferroelectricity.

\subsection*{Retention and Endurance Characteristics}

\begin{figure}
 \centering\includegraphics[width=\linewidth]{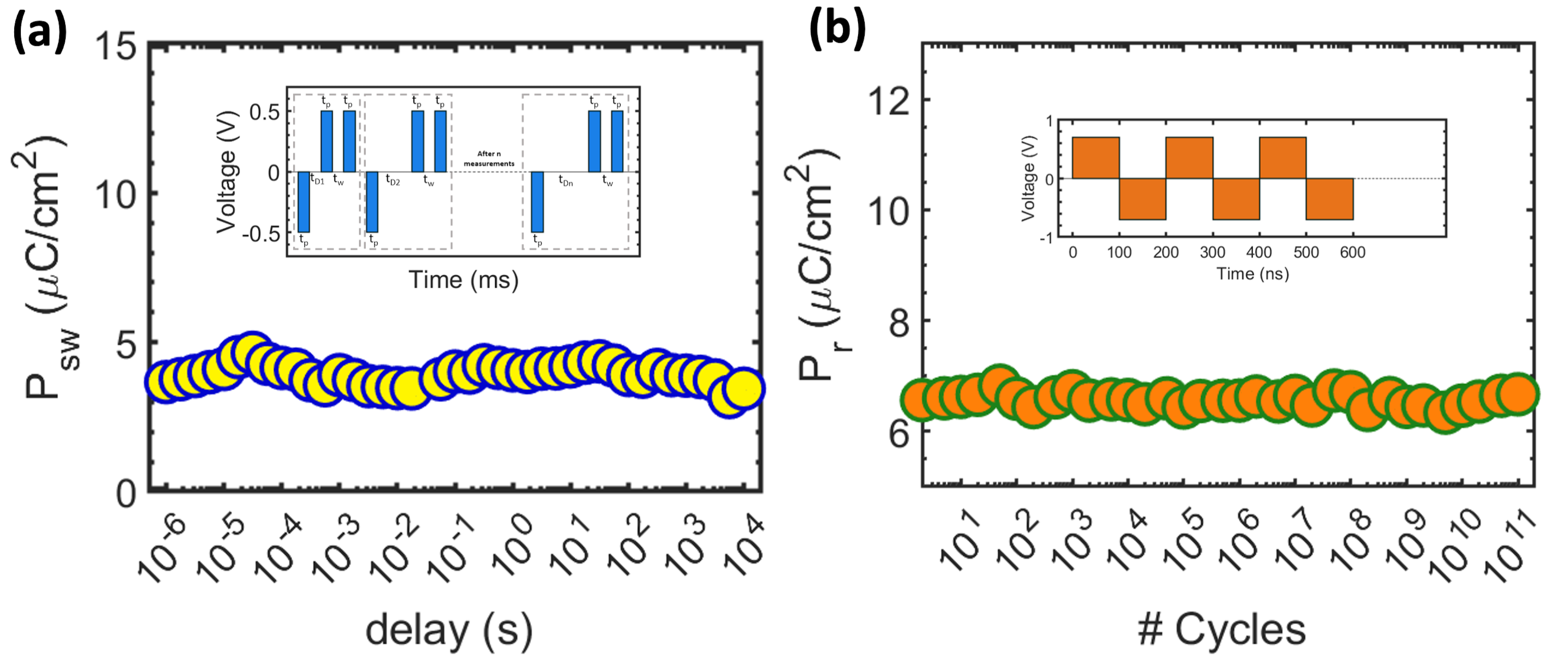}
   \captionof{figure}{(a)Retention characteristics in 20nm BTO film measured till $10^4$ seconds. The applied waveform is shown in the inset of the figure. The pulse width t$_p$ and the wait time between two positive pulses t$_w$ was 100$\mu$s. The Delay between the first and second pulse is increased after each measurement and is represented as t$_{Dn}$ for the n$^{th}$ measurement. The pulse Amplitude was 0.5 V. (b) Endurance characteristics in the same 20 nm BTO film. A continuous square pulse of amplitude 0.7V and pulse width 100ns is applied.}
 \label{fig:Ret}
\end{figure}

Having established the conditions for low leakage currents and robust switching, the retention and endurance characteristics of the 20 nm BTO film grown with the lower SRO plume KEs were examined (spotsizes of 2.92 mm$^2$ and 3.29 mm$^2$ ). A lack of good retention properties in perovskite ferroelectrics has been one of the important reasons in impeding their usage in low voltage switching devices where in their non-volatility is of utmost importance\cite{wan2019nonvolatile}. Typical retention times reported in perovskite ferroelectrics have varied from a few hours\cite{mathews1997ferroelectric} to a few days\cite{watanabe1995epitaxial,aizawa2004impact,hoffman2011device}. However, they have all been reported on films at least a few hundred nanometers thick. The thinner versions ($<$30 nm) have been shown to lose more than 50 percent of their spontaneous polarization within a few seconds\cite{kim2005critical,kim2005polarization,jo2006polarization}. Retention times have been improved by inserting thin layers of high k buffers, thereby reducing leakage currents\cite{aizawa2004impact,takahashi2005thirty}. But they always lead to a substantial increase in the switching voltage. Figure \ref{fig:Ret} shows the retention characteristics of one of the optimized 20 nm BTO films from this study (SRO spotsize of 2.92 mm$^2$ and BTO spotsize of 3.65 mm$^2$). The film shows robust retention with no decay in the spontaneous polarization at least till $10^4$ seconds. It can be seen from figure S18 that in comparison to other oxide ferroelectrics, the films from this work have exceptional retention characteristics. Simultaneously, the films grown in this kinetically controlled regime also showed no drop in remnant polarization up to atleast $1$x$10^{11}$ cycles. This endurance performance also stands out in the thin limit of perovskite ferroelectrics. Endurance and Retention characteristics of a few other heterostructures deposited in the kinetically-controlled regime are shown in figure S19.

\section*{Summary}

In summary, we have demonstrated that by controlling the energies of the plasma plume in PLD, we can achieve sustained ferroelectricity with outstanding retention and endurance properties in ultrathin films of epitaxial BTO capacitors. This control leads to improved bulk and interfacial characteristics. Further, tailoring the non-equilibrium characteristics of the adatoms significantly expands the ``growth window'' for ferroelectric ultrathin films. The results in this study elucidate the nuances and complexities of unexplored regimes in non-equilibrium vapor phase deposition techniques potentially expanding the prospects of large area growth of electronic materials such as ferroelectric thin films for low power electronic applications.

\section*{Methods}
\subsection*{Thin Film Growth}

The epitaxial BTO thin films were deposited on single crystal GdScO$_3$ (110) substrates using Pulsed Laser Deposition (PLD). The BTO is sandwiched between two 20 nm epitaxial SrRuO$_3$ (SRO) layers. For the optimized condition, the BTO films were deposited at a laser fluence of 2.13 J/cm$^2$, at a repetition rate of 2 Hz and at an oxygen partial pressure of 1.1 x 10$^{-1}$ mbar. The corresponding optimized laser fluence, repetition rate and oxygen partial pressure for the deposition of SRO films were 1.4 J/cm$^2$, 15 Hz and 3.4 x 10$^{-2}$ mbar respectively. Both the films were grown at 675$^{\circ}$C. The target-substrate distance used for all layers was 75 mm. The films were continuously monitored \emph{in situ} via Reflection High Energy Electron Diffraction. Subsequently, the films were cooled to room temperature at 3$^{\circ}$C/min under 200 mbar oxygen pressure. 

\subsection*{Structural Characterization}

\subsubsection*{X-Ray Diffraction and reflectivity}

The High Resolution thin film XRD scans in the out of plane and off axis geometry were carried out on a Bruker D8 Advance diffractometer using a Ge (004) two bounce monochromator at Cu K$\alpha _{1}$ ($\lambda$=1.5406 $\si{\angstrom}$) radiation at room temperature. The reciprocal space mapping (RSM) was also carried out at room temperature with Cu K$\alpha _{1}$ radiation by using a Ge(004) 2-bounce monochromator on a Rigaku SmartLab diffractometer with a 9 kW Cu rotating anode. The thickness of each of three layers (bottom SRO, BTO and top SRO) were verified by observing the RHEED oscillations, fitting the Pendellösung fringes in the out of plane XRD and also by fitting the thicknesses of the films through X-Ray Reflectivity (XRR) measurements. The thicknesses were consistent from these three independent measurements. The reflectivity measurements were done on the same diffractometer in a parallel beam geometry using a Göbel (parabolic) mirror set up.

\subsubsection*{Electron Microscopy}

The cross-sectional STEM sample of the SRO/BTO/SRO heterostructure was prepared using a Thermo Scientific Helios G4 PFIB UXe Dual Beam system. Standard \textit{in situ} lift-out was employed during the preparation of the TEM lamella. The sample surface was protected from ion-beam damage by depositing carbon and tungsten protective layers. The TEM lamella was first milled to approximately 140 nm with a 30 kV ion beam, followed by gentler milling at 5 kV to a thickness of approximately 60 nm. A final polishing at 2 kV was performed to minimize surface damage residuals.

HAADF-STEM images were recorded using a probe-corrected Thermo Fisher Scientific Spectra 200 (operated at 200 kV) microscope, equipped with a fifth-order aberration corrector and a X-CFEG cold field emission electron gun. A probe semi-convergence angle of 25 mrad and a current of 100 pA were used during recording. The collection angles were set with inner and outer cutoffs at 53 mrad and 200 mrad, respectively. For each image, multiple frames (over 40) were captured with a single dwell time of 250 ns.

\subsection*{Ferroelectric measurements}

 The epitaxial top layer (SRO) was fabricated into circular electrodes of diameters ranging from 40 $\mu$m to 350 $\mu$m by photolithography and wet etching using 0.4 M NaIO$_4$ solution \cite{weber2013variable}. Electrical characterization was performed on metal-ferroelectric-metal capacitors. P-V, PUND, I-V, retention and fatigue measurements were done using a Keysight B1500A parameter analyzer equipped with WGFMUs. The C-V measurements were performed using a Keysight E4990A Impedance analyzer. The P-V and C-V measurements were performed at 1KHz unless otherwise mentioned. 

 \section*{Acknowledgements}
 This work was supported by the Intel COFEEE program. The thin film growth equipment used in this program was supported in part by the AFOSR DURIP award FA9550-22-1-0117. The Authors acknowledge the use of facilities at the Core Center for Excellence in Nano Imaging at the University of Southern California where in the XRD, XRR and STEM measurements were conducted.

 \section*{Author Contributions}

H.K. and J.R. conceived and designed the experiments on kinetic control through the spotsize. H.K. grew the films using PLD, structurally characterized them using XRD and XRR and fabricated the contacts on the films. P.V.R. and A.I.K. designed the extensive ferroelectric characterization measurements. P.V.R. performed the various electrical characterization measurements and was assisted by T.S.. T.L. and Y.S. performed the STEM measurements and T.L. performed the analysis for the same. M.S. was involved in the structural characterization measurements of these films and in discussions on the technical details pertaining to the growth of the films. H.C. provided technical support on the fabrication and preliminary electrical measurements. P.B. and I.T. performed electrical measurements and Diffraction measurements respectively on some representative samples. P.B., I.T., A.S.G., R.S. and I.A.Y. contributed to the analysis and understanding of the data. H.K. and J.R. wrote the manuscript with inputs from other authors. 

\newpage
\setcounter{figure}{0}
\Huge 
\begin{center}
    \maketitle{Supporting Information}
\end{center}

\normalsize

\onehalfspacing

\section{RHEED - \textit{in situ} characterization}
\renewcommand\thefigure{S\arabic{figure}}

\begin{figure}[h]
 \centering\includegraphics[width=\linewidth]{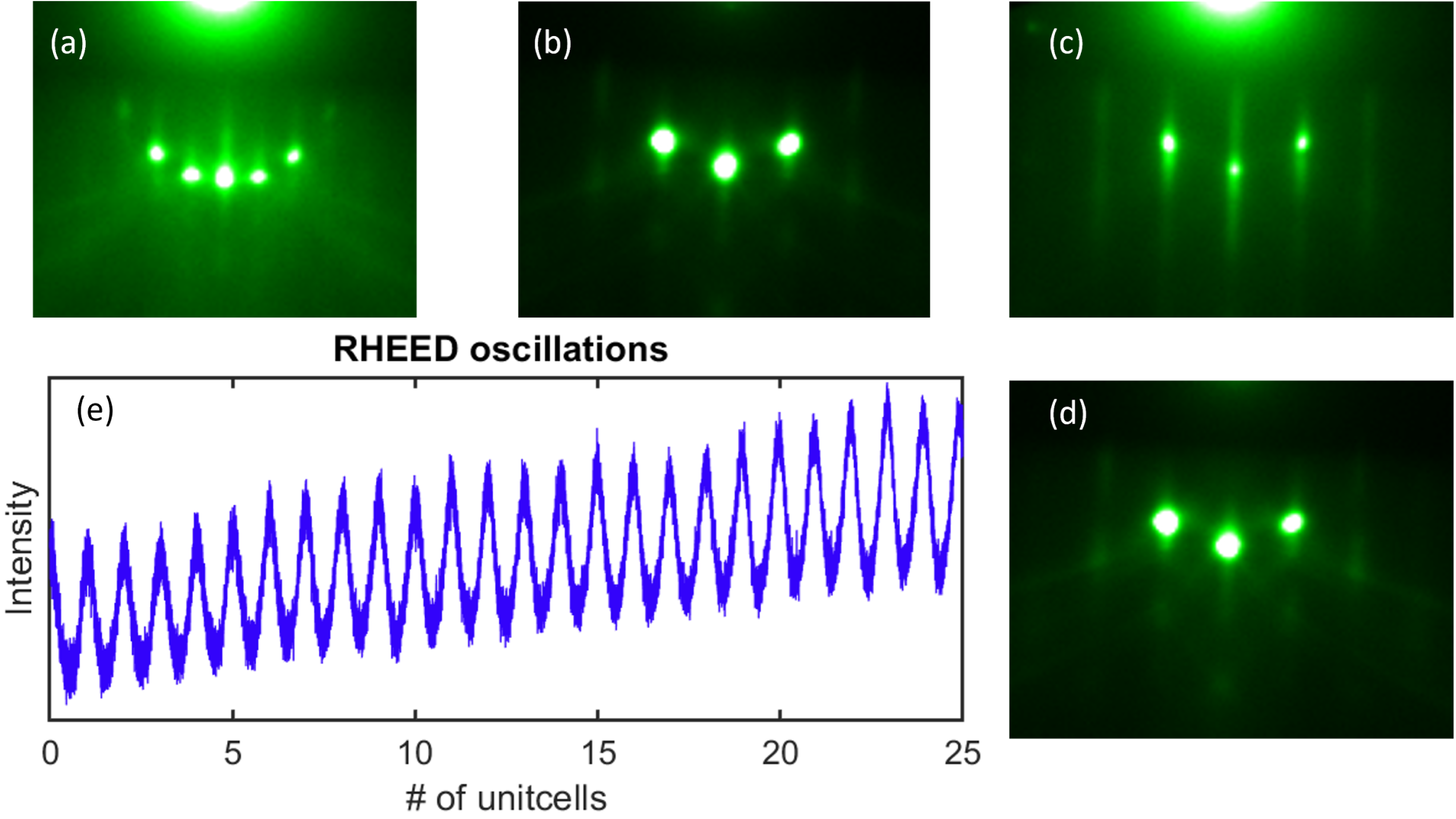}
 \captionof{figure}{ (a) GdScO$_{3}$ (110) substrate before growth (b) RHEED pattern of SrRuO$_{3}$ (SRO) bottom electrode (c) RHEED pattern of BaTiO$_{3}$ (BTO) (d) RHEED pattern of SRO top electrode at the end of growth (e) RHEED oscillation during BTO growth.}
 \label{fig:RHEED}
\end{figure}

Reflection High Energy Electron Diffraction was used \textit{in situ} to monitor the surface of the substrates and films during growth. Figures \ref{fig:RHEED} (a)-(d) show the diffraction patterns of the substrate surface and the surface after the deposition of each layer. The Diffraction patterns for the optimized conditions show a smooth 2D surface as indicated. Intensity oscillations were observed throughout the growth of BTO (Figure \ref{fig:RHEED} (e)).

\vspace{6pt}

\section{Growth Window and growth modes for 100 nm BTO films}

\begin{figure}
 \centering\includegraphics[width=0.8\linewidth]{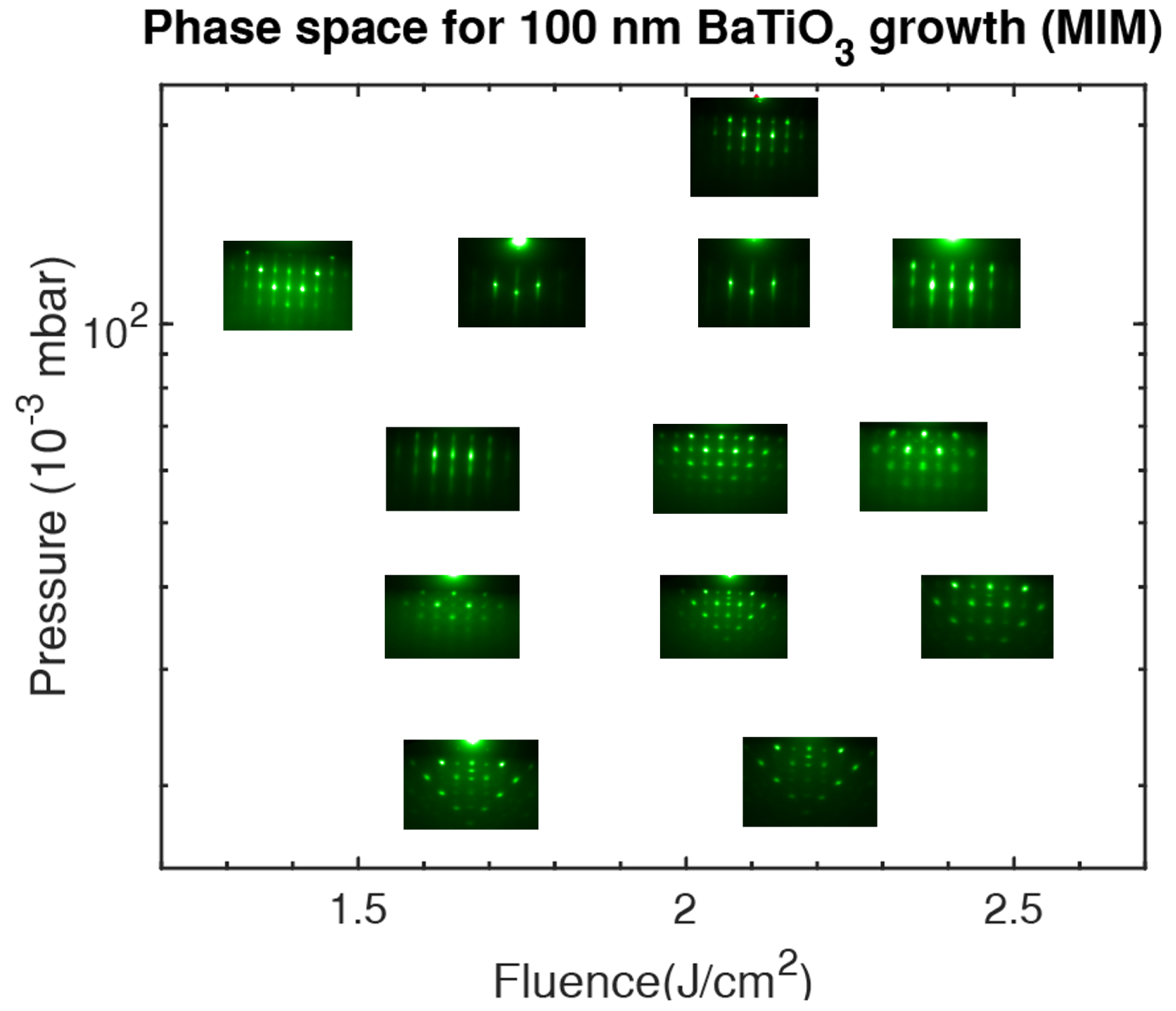}
 \captionof{figure}{The RHEED patterns for 100 nm BTO films grown on 20nm SRO/GSO heterostructures as a function of oxygen pressure and laser fluence. The growth conditions for SRO were held constant. }
 \label{fig:100nmRHEED}
\end{figure}

\begin{figure}
 \centering\includegraphics[width=\linewidth]{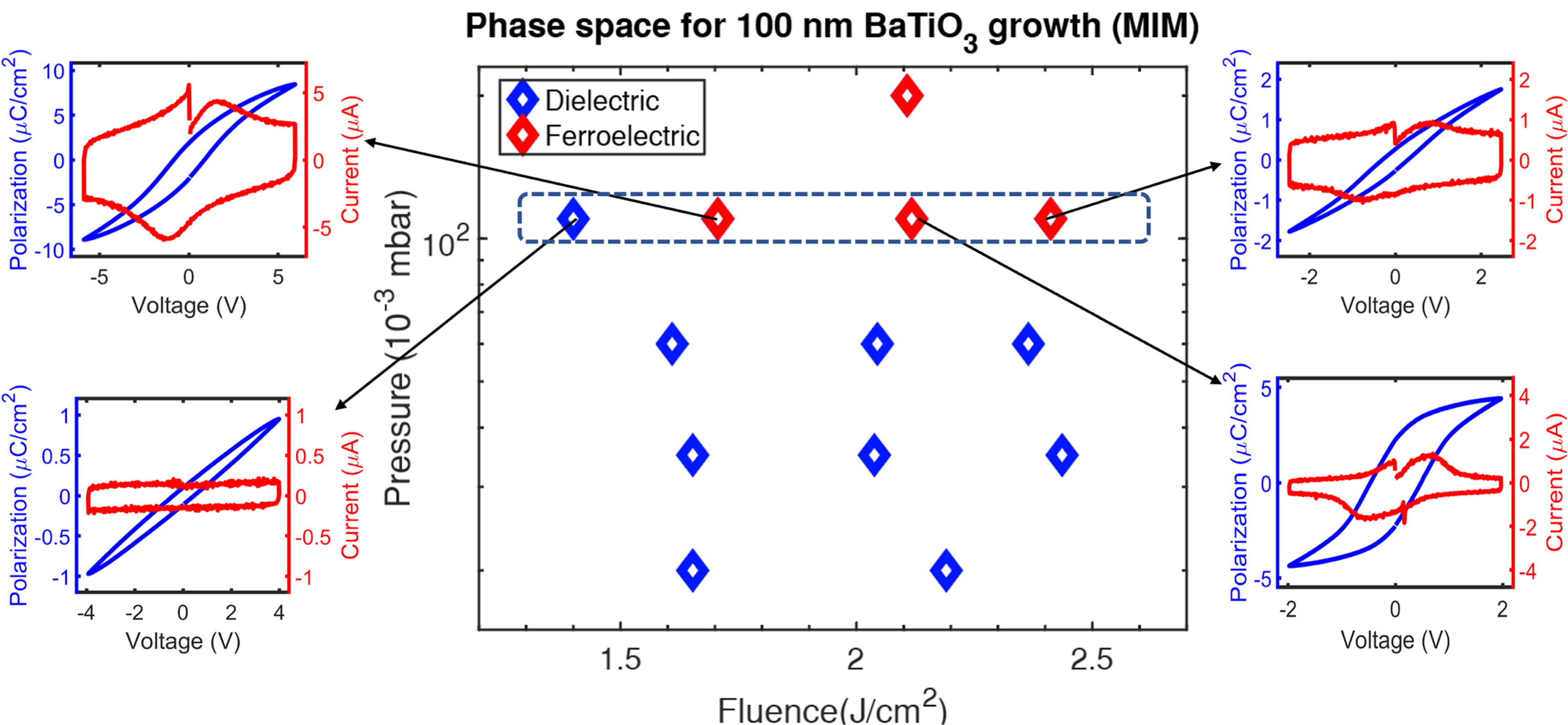}
 \captionof{figure}{The evolution of dielectric properties of 100nm BTO films as a function of oxygen pressure and laser fluence.}
 \label{fig:100nmferrophasespace}
\end{figure}

The initial optimization of growth parameters was performed on 100 nm BTO/ 20nm SRO/GdScO$_{3}$ (GSO) heterostructures by varying the laser fluence and background gas pressure for the BTO layer. The growth conditions for the SRO layer was fixed for all these films. A spotsize of 2.92 mm$^2$ was used for the BTO and 3.65 mm$^2$ for the SRO layers. An amorphous SRO top electrode was fabricated post-growth for these 100 nm films. Figure \ref{fig:100nmRHEED} shows the RHEED patterns for the 100 nm BTO films and it can be seen that the films have a 2D or close to 2D pattern at 0.11 mbar oxygen pressure (growth pressure for  BTO layers only). Also, most of the films grown at this pressure were ferroelectric (figure \ref{fig:100nmferrophasespace}). Hence this pressure was fixed for BTO in the further growths at lower thicknesses and the variations in spotsizes.

\par

\vspace{6pt}

\par

\section{X-ray Diffraction as a function of thickness}

\par
\begin{figure}[h]
 \centering\includegraphics[width=\linewidth]{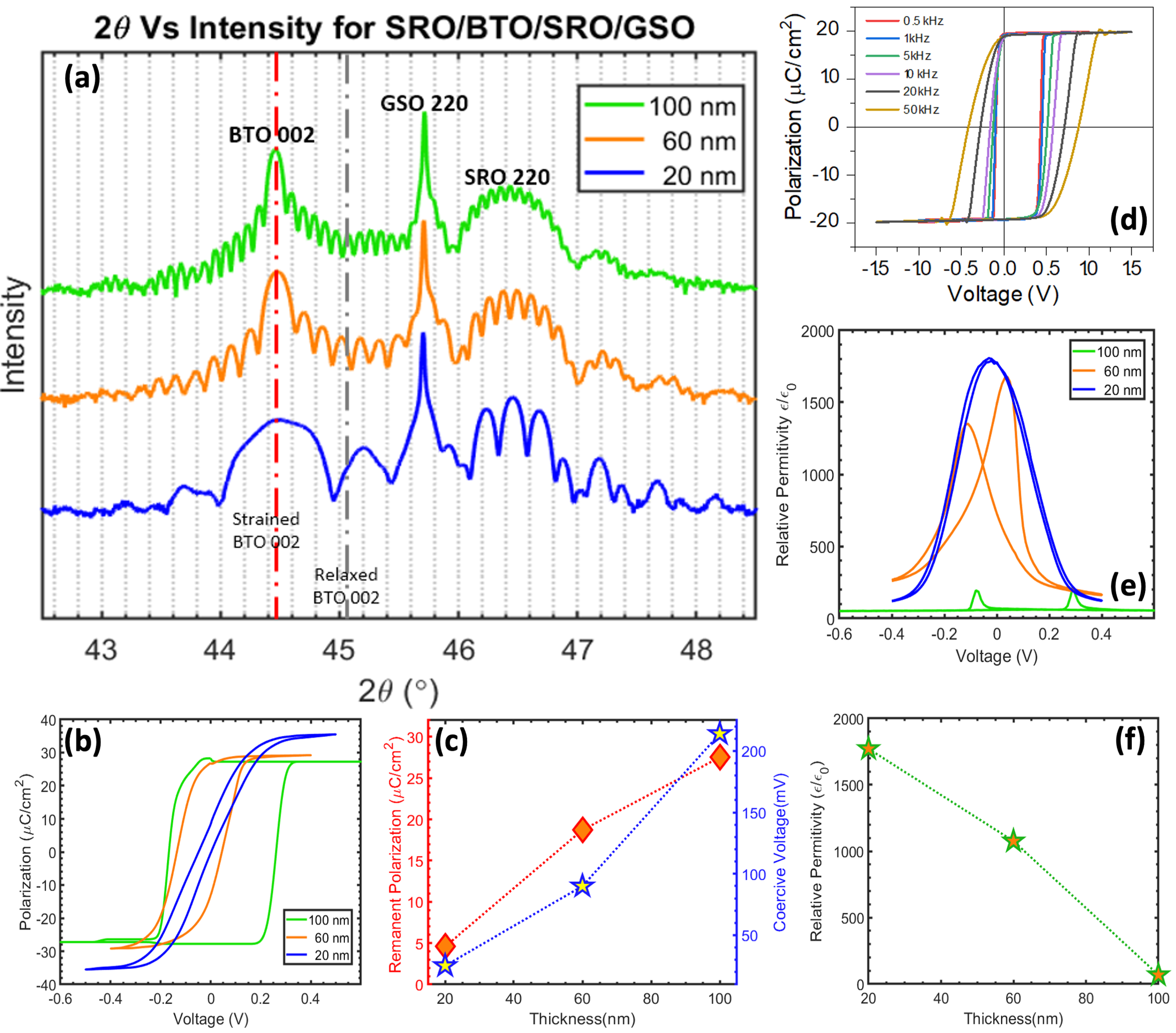}
 \captionof{figure}{(a)Thickness variation for the out of plane thin film X-ray Diffraction profiles of SRO/BTO/SRO/GSO heterostructures. (b) Polarization-Voltage curves as a function of BTO thickness at 1 KHz. (c) Variation of remanent Polarization and Coercive voltage with thickness. (d) Frequency-dependant Polarization Polarization-Voltage curves for the SRO/100 nm BTO/SRO/GSO heterostructure. (e) Capacitance-Voltage curves as a function of BTO thickness at 1KHz. (f) Variation of Relative Permittivity with thickness }
 \label{fig:PrEcVcErtrends}
\end{figure}

\vspace{6pt}

Figure \ref{fig:PrEcVcErtrends}(a) shows the out-of-plane X Ray Diffraction profiles for the SRO/BTO/SRO heterostructures as a function of BTO thicknesses for the optimized BTO condition - 2.13 J/cm$^2$ and 1.1 x 10$^{-1}$ mbar. It can be seen that BTO films of all thicknesses are textured along 002 (\textit{c}-axis) in the out of plane direction.

\section{Trends in ferroelectric properties}

Figure \ref{fig:PrEcVcErtrends}(b) shows the polarization-voltage hysteresis loops for BTO films of different thicknesses. All films show robust ferroelectricity with a saturation polarization of about 30 $\mu$C/cm$^2$. The remanent polarization decreases with the decrease in thickness of films (Figure \ref{fig:PrEcVcErtrends}(c)).The slanted and relaxor-like nature of the P-V characteristics gradually becomes prominent for thinner films, especially for the 20 nm film. In a capacitor geometry, when the free charges on the conducting electrodes (SRO in our case) are distributed over a finite length, the bound charges in the ferroelectric aren't compensated completely\cite{mehta1973depolarization}. This imperfect screening would result in a depolarization field that opposes the spontaneous polarization of the ferroelectric layer. This depolarization effect increases as the charge distribution in the metal electrode deviates from an ideal sheet charge configuration and also as the volume of the bulk of the ferroelectric diminishes with respect to that of the interface. This depolarization effect may lead to multi-domain formation, which is manifested as relaxor-like behavior in P-V characteristics.

The increase in depolarization field with reduction in thickness is also very evident from the C-V characteristics (Figure \ref{fig:PrEcVcErtrends}(e)). The depolarization field is manifested as an effective reduction in the Curie temperature of the ferroelectric layer. This results in an increase in the dielectric constant(Figure \ref{fig:PrEcVcErtrends}(f)), which are often higher at temperatures closer to the ferroelectric transition. Here, we observe that the dielectric constant at the zero bias are higher for the thinner films, which is consistent with an effective reduction of the Curie temperature.

\section{PUND measurements}

\par
\begin{figure}[h]
 \centering\includegraphics[width=\linewidth]{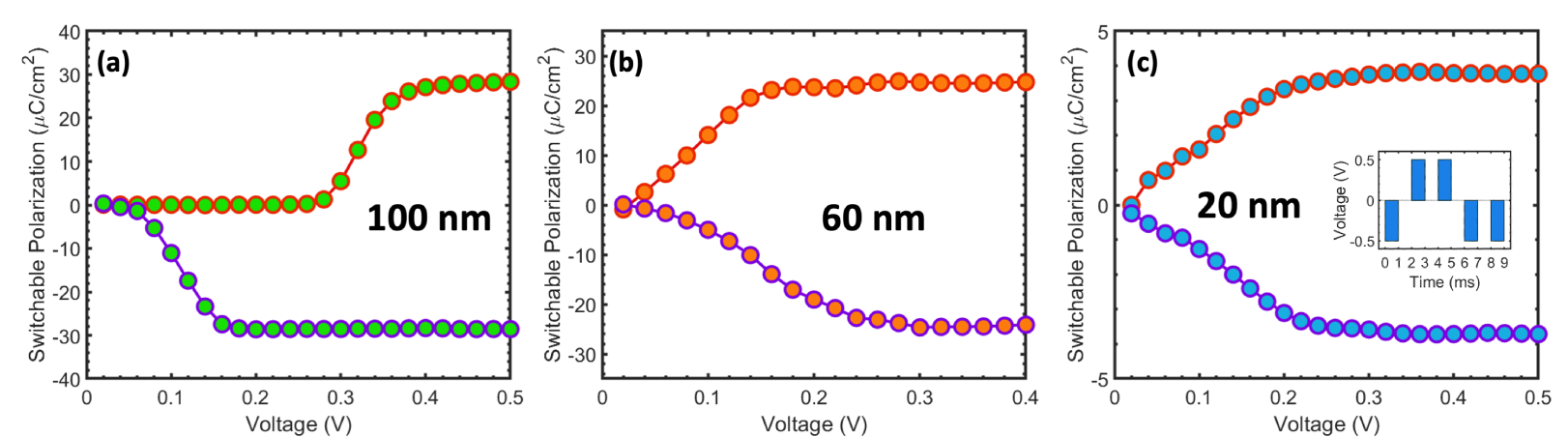}
 \captionof{figure}{ Switching charge Vs voltage during PUND measurements for (a) 100 nm, (b) 60 nm and (c) 20 nm BTO films. Pulse sequence applied is shown in the inset of (c). }
 \label{fig:PUND}
\end{figure}

\vspace{6pt}

Considering the very slanted P-V loop in addition to the barely distinguishable double peak in C-V characteristics of the 20 nm BTO, it is critical to verify its ferroelectric nature. In fact, a hysteresis loop doesn’t guarantee ferroelectricity as it accounts for the total switched charge and not just the one from the reversal in direction of the spontaneous polarization alone. A leaky, non-linear dielectric would also show a hysteresis loop that contains a time dependent leakage component. To isolate the ferroelectric component, we used a sequence of voltage pulses and measured the current transients - also known as the PUND method. Figure \ref{fig:PUND} shows the switched polarization for all thicknesses as a function of voltage and a non-zero switched polarization confirms the ferroelectricity in all these films. All films reported as ferroelectric in this work show a non-zero switching charge via the PUND method.

\section{Laser Spotsize variation in 20nm SRO/20nm BTO/20nm SRO heterostructures.}

\begin{figure}[h]
 \centering\includegraphics[width=0.7\linewidth]{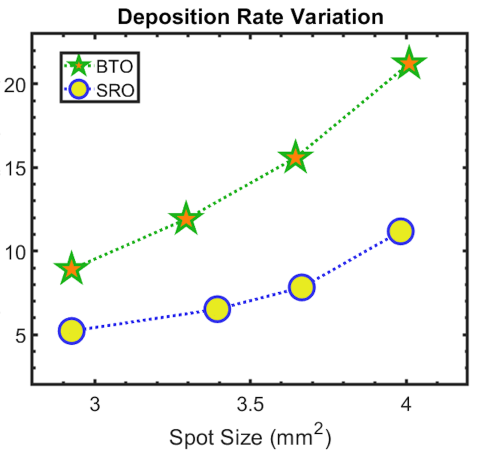}
 \captionof{figure}{Deposition rates of BTO and SRO as a function of the laser spotsize}
 \label{fig:deprate}
\end{figure}

Figure \ref{fig:deprate} shows the change in deposition rates for both BTO and SRO as a function of the laser spotsize. It can be seen that as the spotsize becomes smaller, the deposition rate correspondingly decreases. The deposition rates were measured by tracking the intensity oscillations in RHEED and verified through X-ray Reflectivity fits for both the materials. 

\vspace{6pt}

\section{Dielectric Properties of 20nm SRO/ 20nm BTO/ 20nm SRO/GSO heterostructures}

The phase angle, leakage currents and the switching charge in PUND for the different spot sizes used for the 20nm SRO/ 20nm BTO/ 20nm SRO/GSO heterostructures are shown in figure \ref{fig:loss,leakage,PUND}

\begin{figure}
 \centering\includegraphics[width=\linewidth]{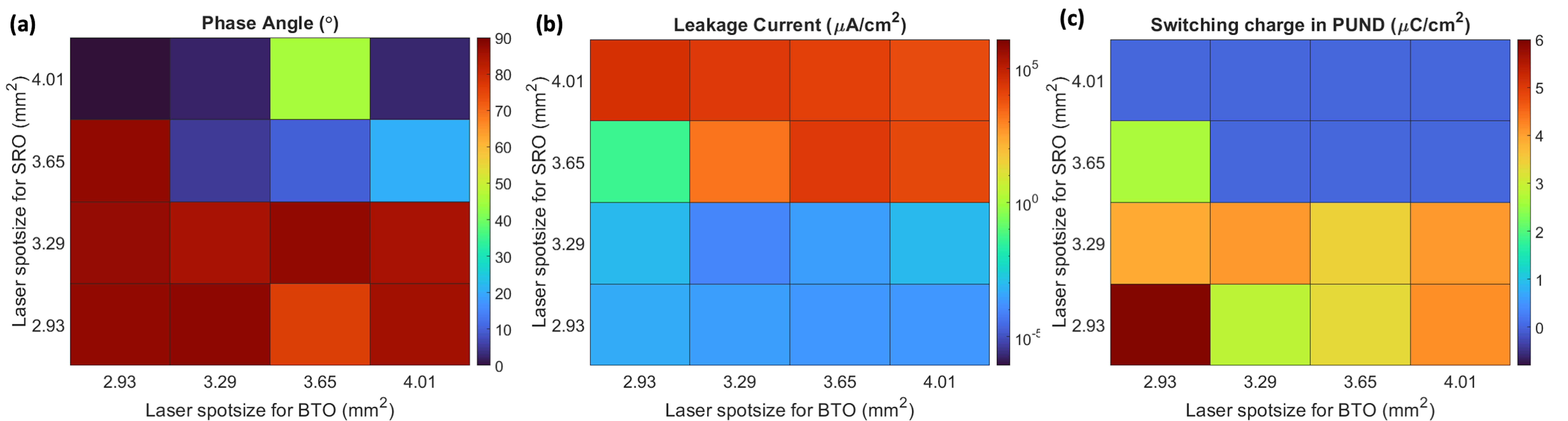}
   \captionof{figure}{(a)Phase angle at 0.5V (b) Leakage currents at 1V and (c) the switching charge in PUND in 20nm SRO/ 20nm BTO/ 20nm SRO/GSO heterostructures as a function of the laser spotsizes.}
 \label{fig:loss,leakage,PUND}
\end{figure}

\vspace{6pt}

\begin{figure}
 \centering\includegraphics[width=0.65\linewidth]{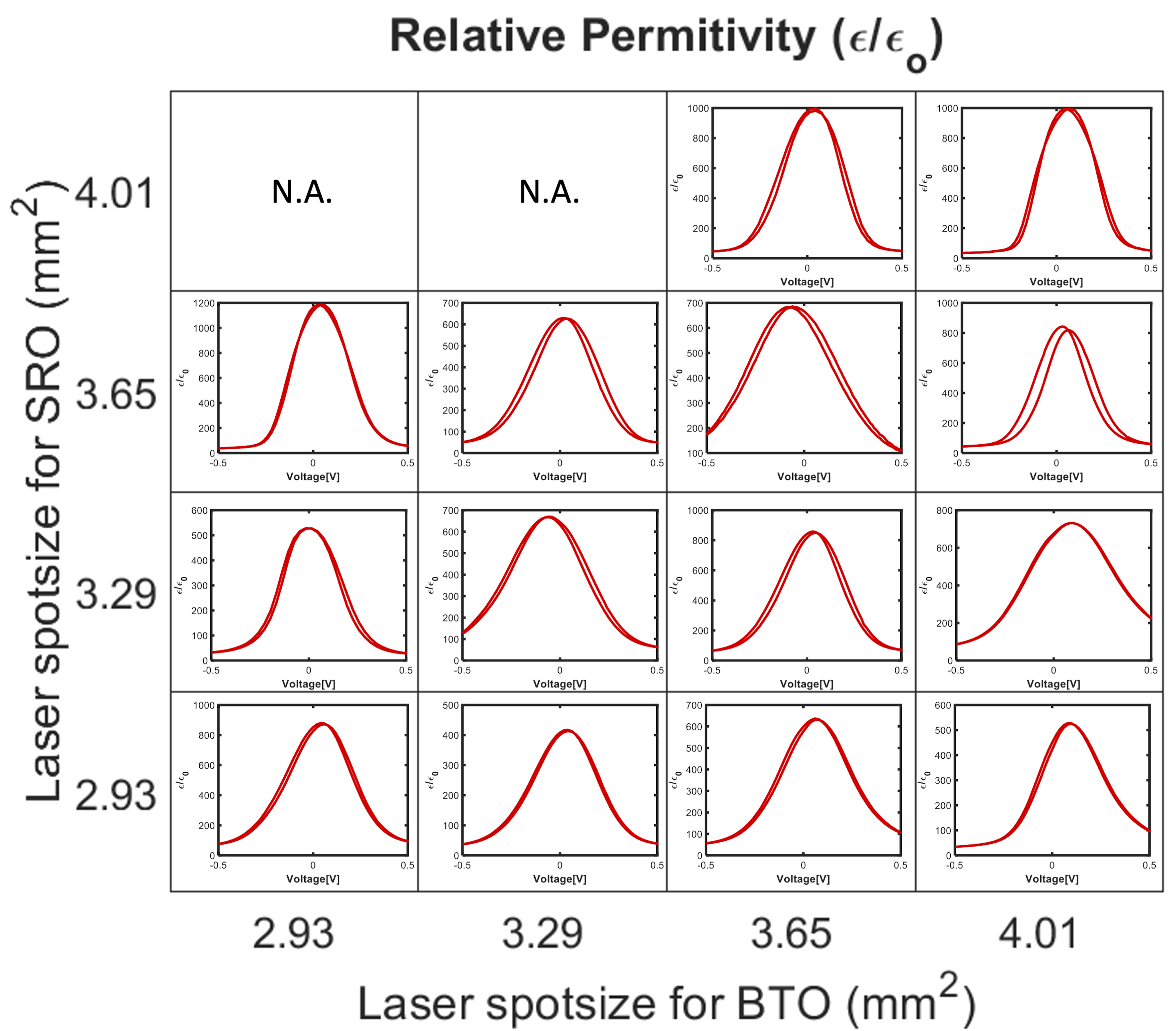}
   \captionof{figure}{Relative permittivity in 20nm SRO/ 20nm BTO/ 20nm SRO/GSO heterostructures as a function of the SRO and BTO laser spotsizes. The heterostructures with the SRO spotsize of 4.01 mm$^2$ and BTO spotsizes of 2.93 and 3.29 mm$^2$ have large leakage currents such that a C-V curve couldn't be extracted.}
 \label{fig:CV_ss_20nm}
\end{figure}

\begin{figure}
 \centering\includegraphics[width=0.65\linewidth]{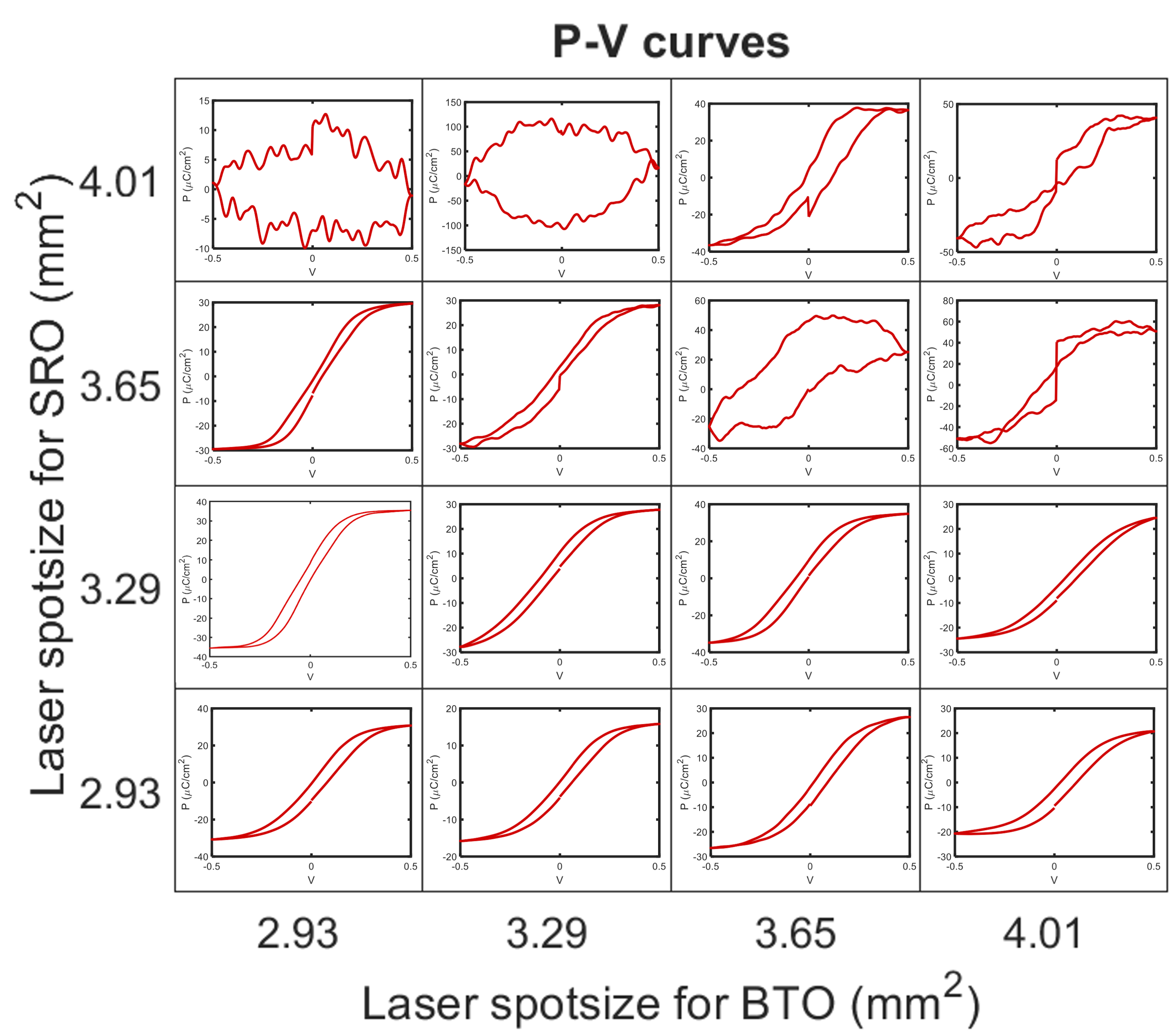}
   \captionof{figure}{Polarization($\mu$C/cm$^2)$-Voltage Curves for the 20nm SRO/ 20nm BTO/ 20nm SRO/GSO heterostructures as a function of the SRO and BTO laser spotsizes.}
 \label{fig:PV_ss_20nm}
\end{figure}

\begin{figure}
 \centering\includegraphics[width=0.65\linewidth]{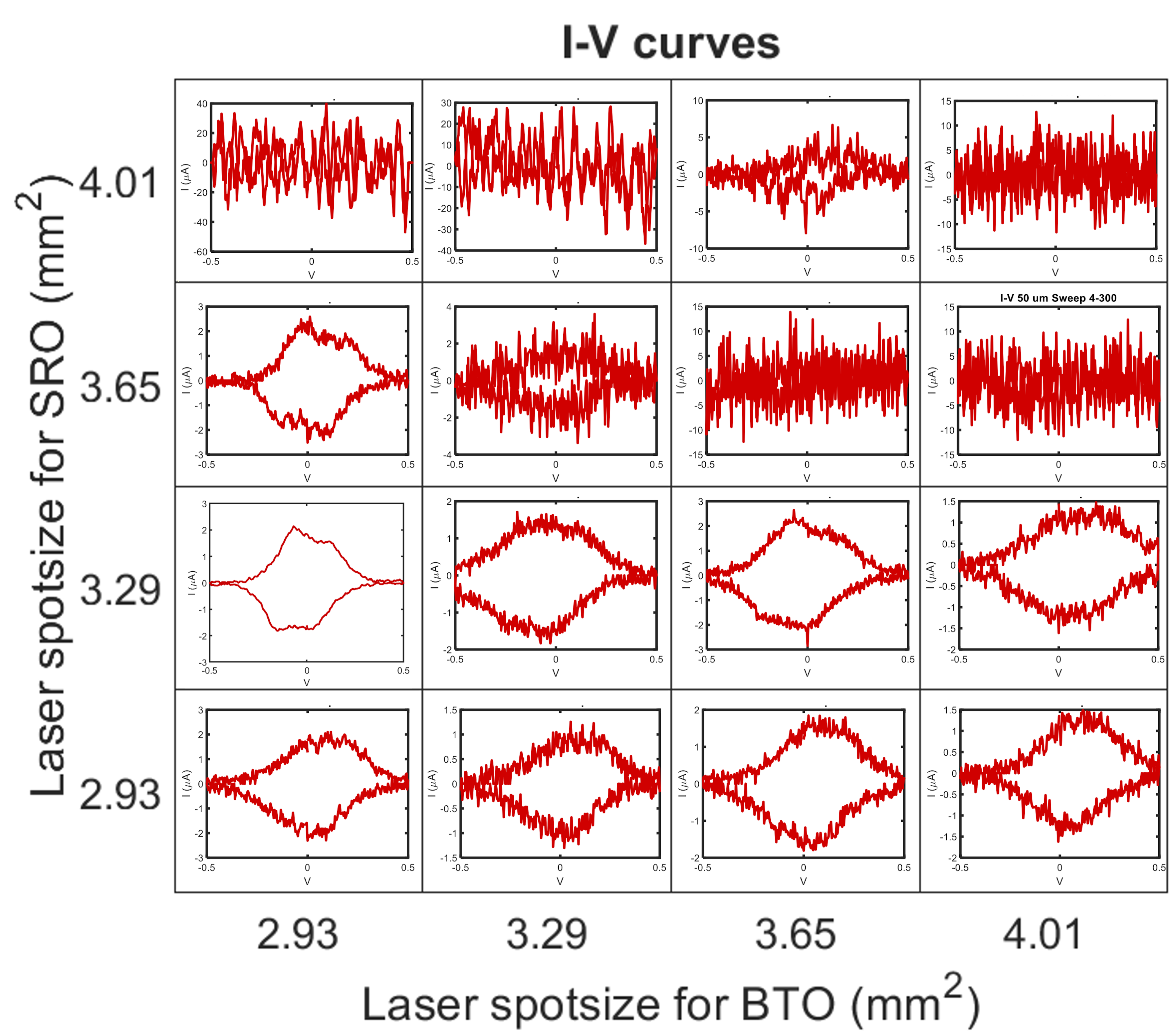}
   \captionof{figure}{Switching Current($\mu$A)-Voltage Curves for the 20nm SRO/ 20nm BTO/ 20nm SRO/GSO heterostructures as a function of the SRO and BTO laser spotsizes.}
 \label{fig:IV_ss_20nm}
\end{figure}

\begin{figure}
 \centering\includegraphics[width=0.65\linewidth]{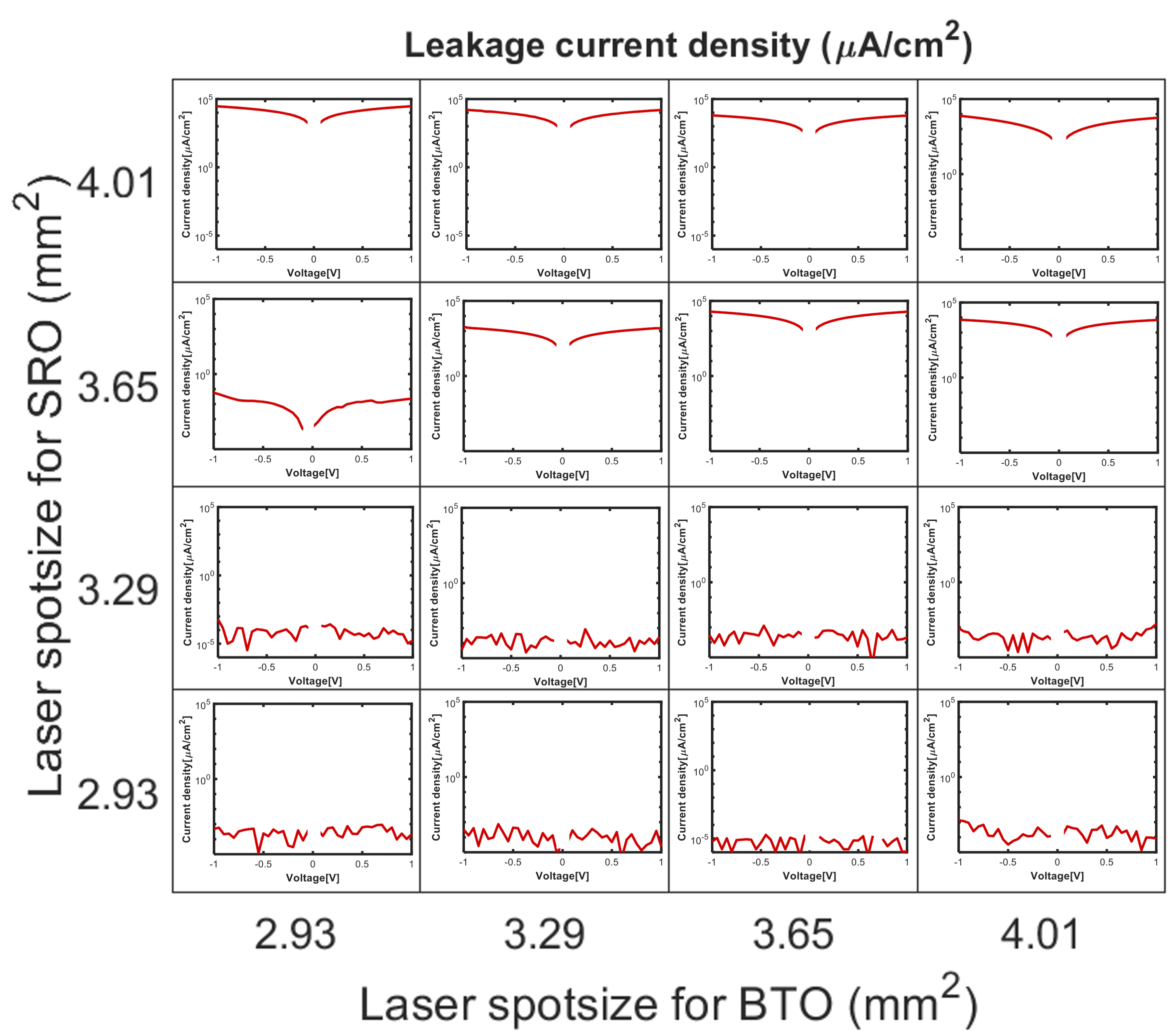}
   \captionof{figure}{Leakage Current Density($\mu$A/cm$^2)$ curves for the 20nm SRO/ 20nm BTO/ 20nm SRO/GSO heterostructures as a function of the SRO and BTO laser spotsizes.}
 \label{fig:leakage_ss_20nm}
\end{figure}

\begin{figure}
 \centering\includegraphics[width=0.7\linewidth]{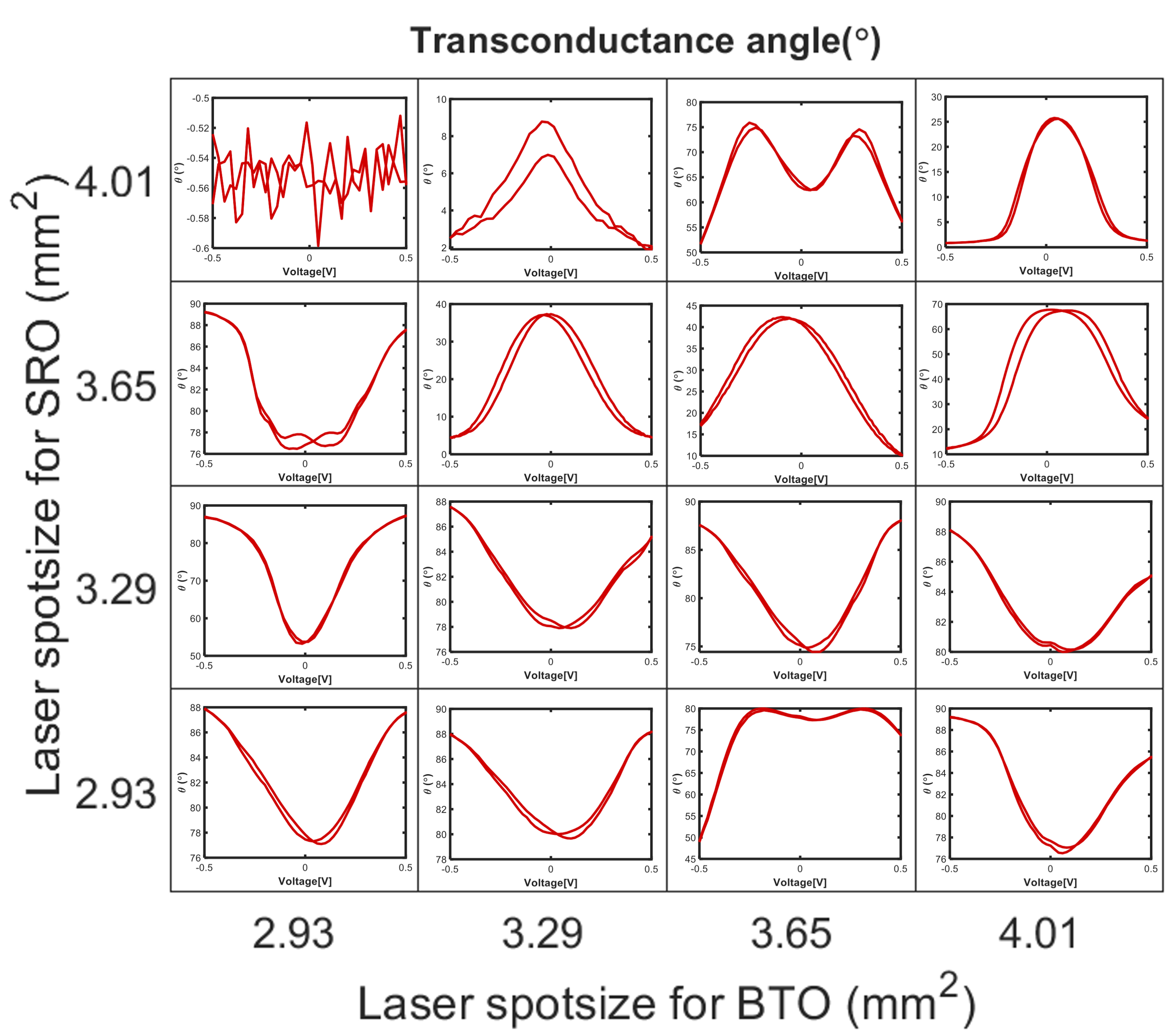}
   \captionof{figure}{The Transconductance angle for the 20nm SRO/ 20nm BTO/ 20nm SRO/GSO heterostructures as a function of the SRO and BTO laser spotsizes. }
 \label{fig:phase_ss_20nm}
\end{figure}

\begin{figure}
 \centering\includegraphics[width=0.7\linewidth]{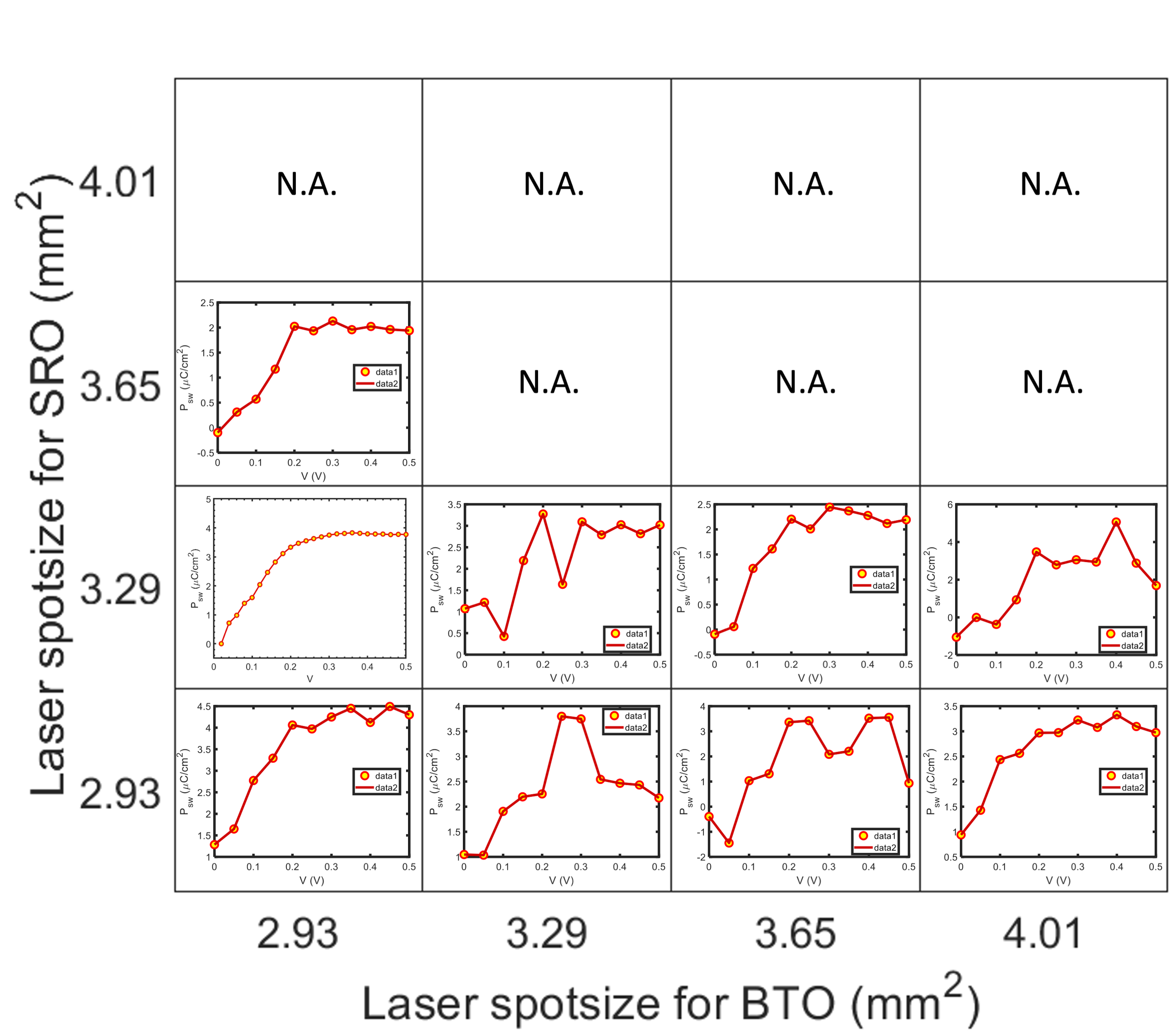}
   \captionof{figure}{The PUND curves for the 20nm SRO/ 20nm BTO/ 20nm SRO/GSO heterostructures as a function of the SRO and BTO laser spotsizes. The applied pulse sequence is the same as the one shown in the inset of figure \ref{fig:PUND}(c)}
 \label{fig:PUND_ss_20nm}
\end{figure}

The individual dielectric properties of the different 20nm SRO/ 20nm BTO/ 20nm SRO/GSO heterostructures in the spotsize phase space, namely the P-V,C-V, switching currents, leakage currents, phase angle and the PUND curves are shown in figures \ref{fig:CV_ss_20nm}, \ref{fig:PV_ss_20nm}, \ref{fig:IV_ss_20nm}, \ref{fig:leakage_ss_20nm},  \ref{fig:phase_ss_20nm} and \ref{fig:PUND_ss_20nm} respectively.

\clearpage

\section{Off Axis Reflections and Reciprocal Space Maps}

\begin{figure}[h]
 \centering\includegraphics[width=0.65\linewidth]{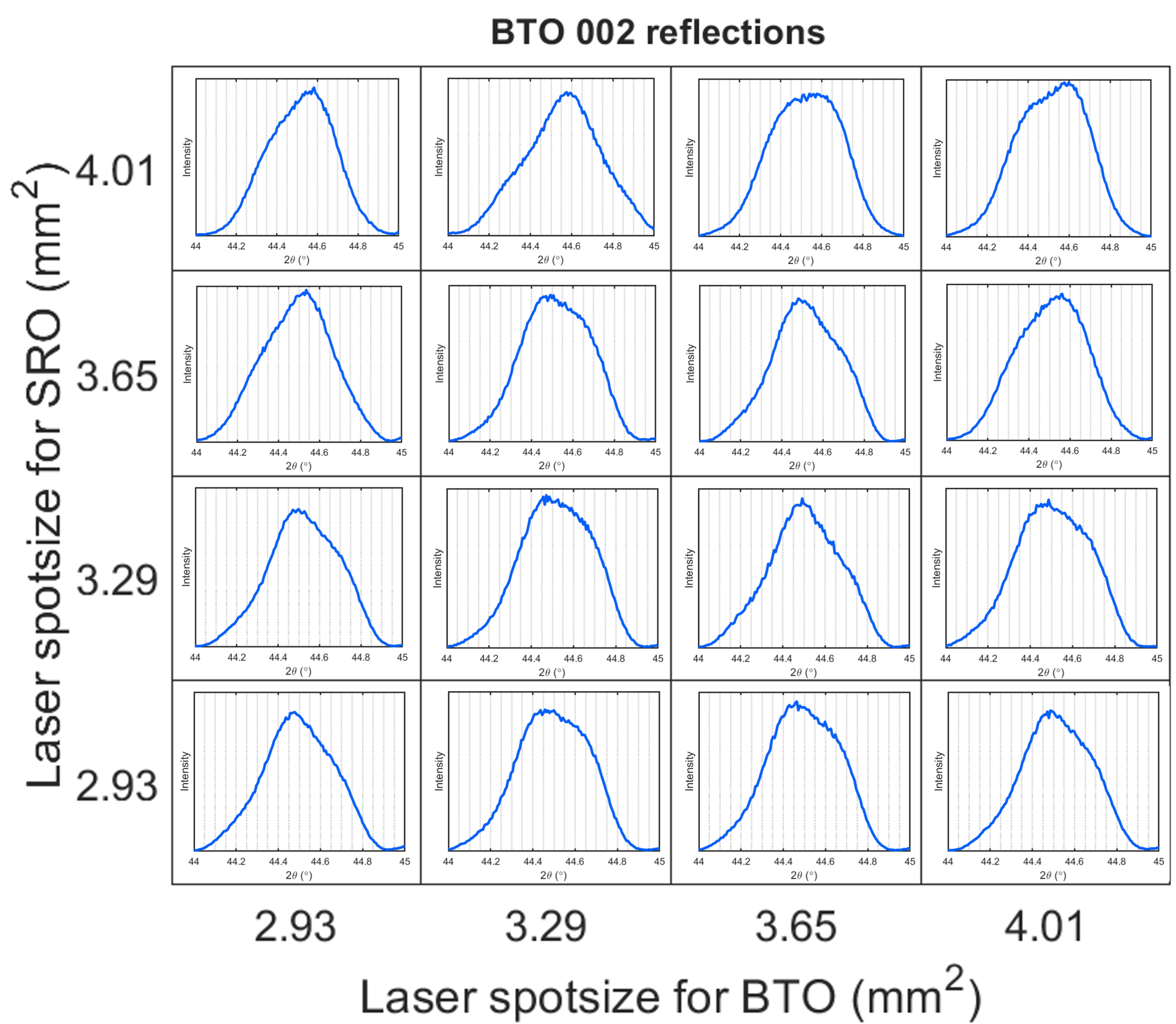}
   \captionof{figure}{BTO 002 reflection from the 20nm SRO/ 20nm BTO/ 20nm SRO/GSO heterostructures as a function of the SRO and BTO laser spotsizes. The intensity is in linear scale.}
 \label{fig:BTO002_20nm}
\end{figure}

\begin{figure}
 \centering\includegraphics[width=0.65\linewidth]{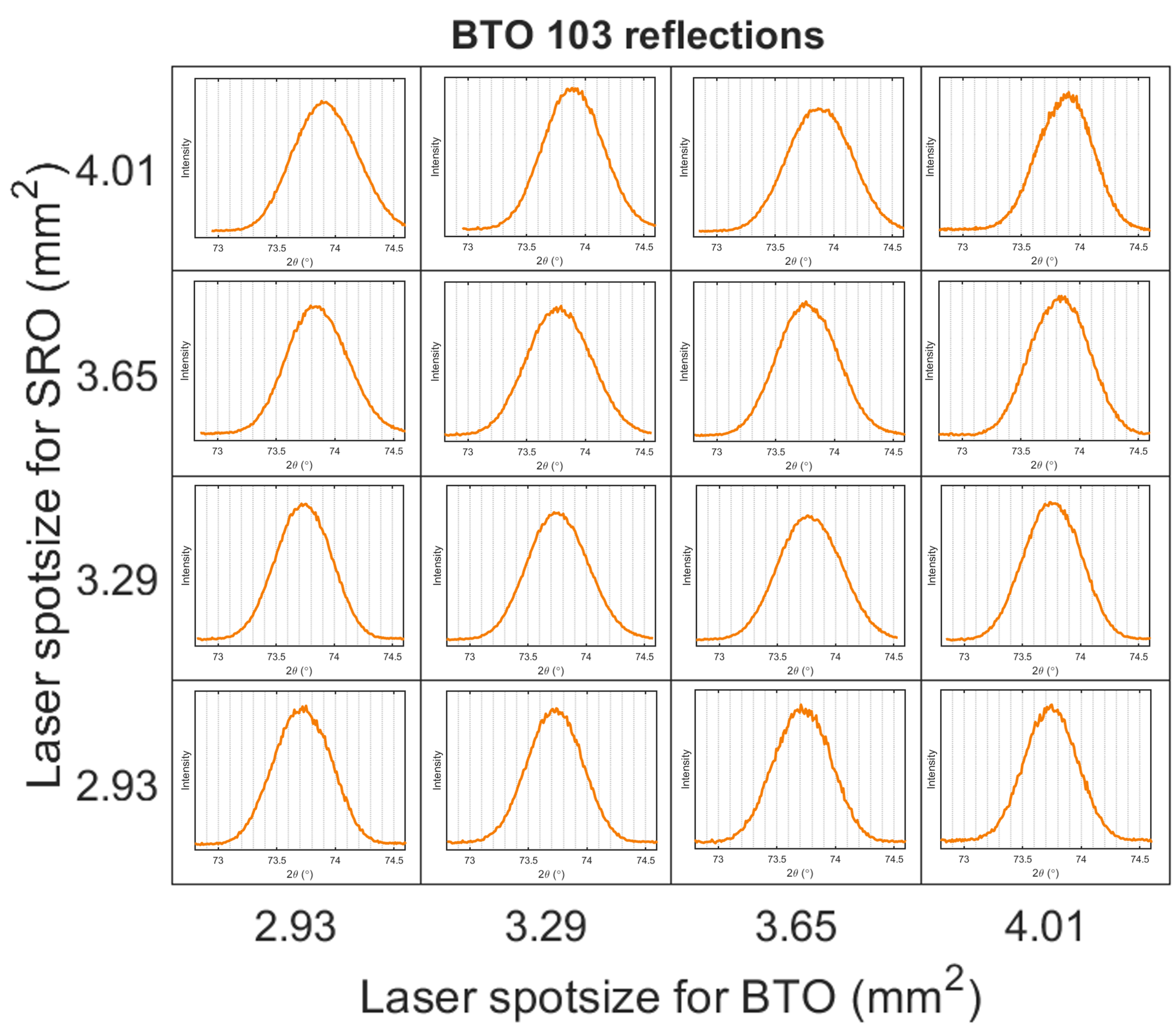}
   \captionof{figure}{BTO 103 reflection from the 20nm SRO/ 20nm BTO/ 20nm SRO/GSO heterostructures as a function of the SRO and BTO laser spotsizes. The intensity is in linear scale.}
 \label{fig:BTO103_20nm}
\end{figure}

\begin{figure}[h]
 \centering\includegraphics[width=0.95\linewidth]{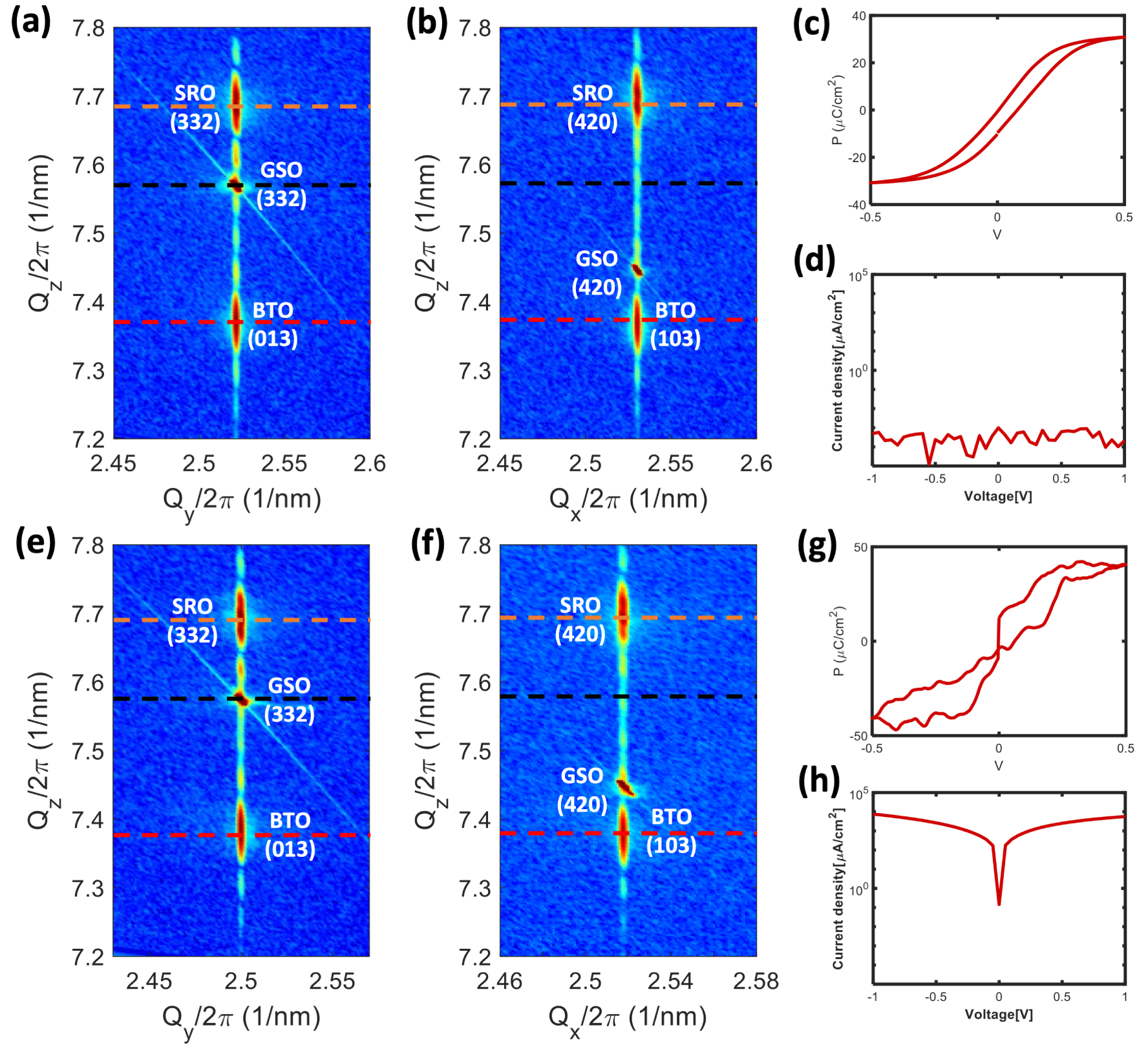}
   \captionof{figure}{ Reciprocal Space Maps obtained around (a) GSO 332 and (b) 420 for a ferroelectric 20nm SRO/ 20nm BTO/ 20nm SRO/GSO heterostructure with a laser spotsize of 2.92 mm$^2$ used for all layers. (c) Corresponding P-V and (d) leakage current density for this heterostructure.
   Reciprocal Space Maps obtained around (e) GSO 332 and (f) 420 for a high leakage 20nm SRO/ 20nm BTO/ 20nm SRO/GSO heterostructure with a laser spotsize of 4.01 mm$^2$ used for all layers. (g) Corresponding P-V and (h) leakage current density for this heterostructure.}
 \label{fig:RSM}
\end{figure}

\vspace{6pt}

Figure \ref{fig:RSM} shows the Reciprocal Space Maps (RSM) obtained for two representative heterostructures. The in-plane lattice parameters of BTO and SRO are coincident with the GSO substrate, indicating that both the BTO and SRO films are coherently strained to the GSO substrate. This is the case for both the high leakage and the ferroelectric film. The films were picked from the two extremes of the phasespaces - the ferroelectric heterostructure was made by using the lowest spotsize of 2.92 mm$^2$ for all layers whereas the high leakage sample was made by using the largest spotsize of 4.01 mm$^2$ for all layers.

\clearpage

\section{Frequency dependant P-V characteristics}

\begin{figure}[h]
 \centering\includegraphics[width=\linewidth]{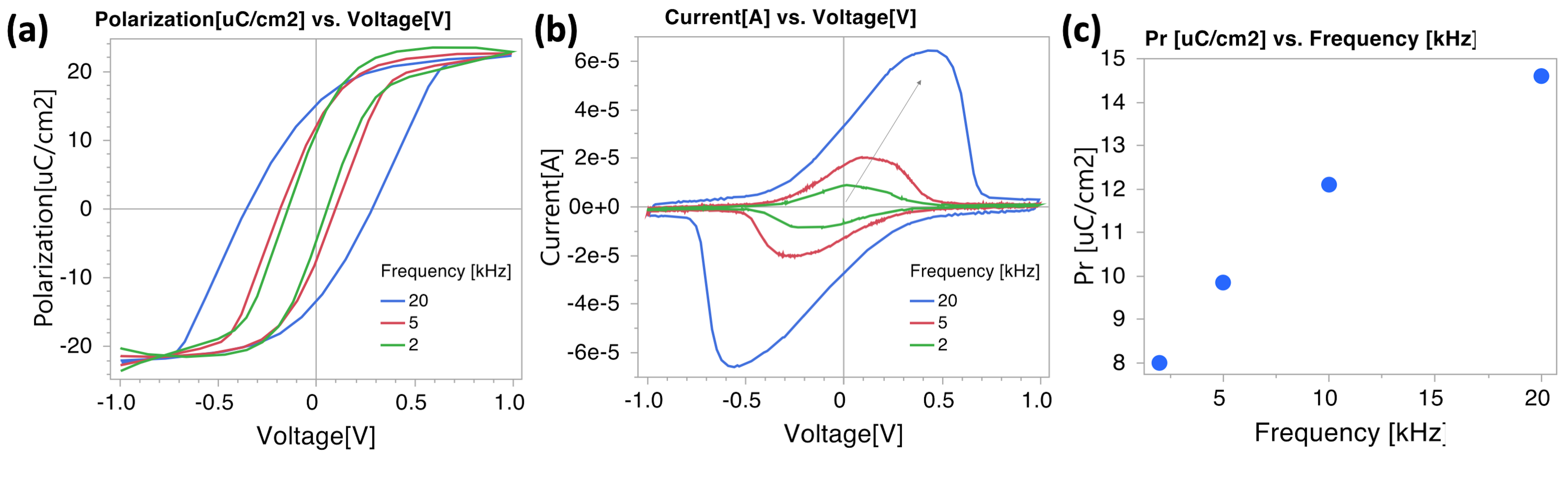}
   \captionof{figure}{Frequency dependency in the (a) P-V loops, (b) switching Currents and (c) Remanent Polarization in 20nm SRO/ 20nm BTO/ 20nm SRO/GSO heterostructure with the lowest BTO and SRO spotsizes (2.92 mm$^2$ in this study)}
 \label{fig:Inteldata}
\end{figure}

\section{Current Status of Retention and Endurance in Ferroelectric Thin Films}

\begin{figure}
 \centering
 \includegraphics[width=\linewidth]{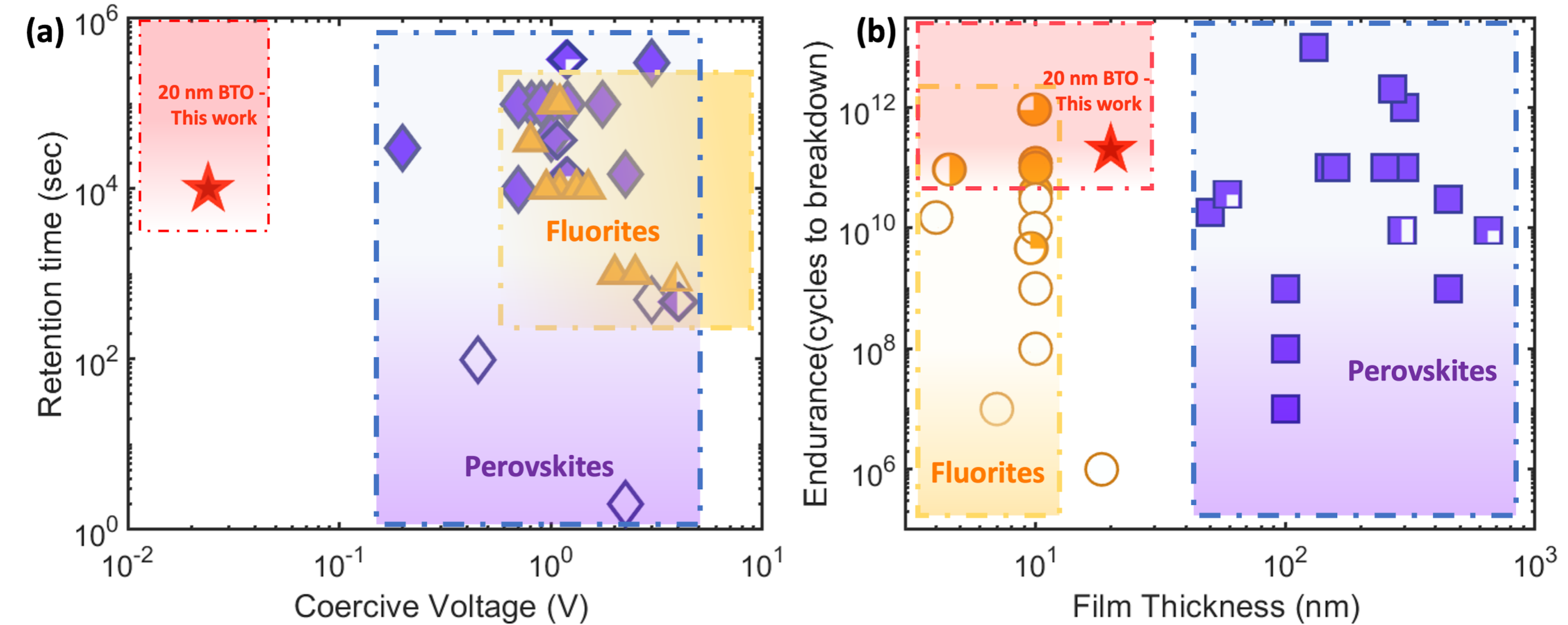}
 \captionof{figure}{(a) Benchmark plot showing retention times for fluorite oxide ferroelectrics (orange triangles) and perovskite oxide ferroelectrics (purple diamonds) against the 20 nm BTO sample from this work. The fraction of the shaded regions indicates the percentage of polarization left at the times indicated. (b) Benchmark plot showing Endurance (number of cycles) for fluorite oxide ferroelectrics (orange circles) and perovskite oxide ferroelectrics (purple squares) against the 20 nm BTO sample from this work. The fraction of the shaded regions indicates the percentage of polarization left after the indicated number of cycles. The references for the benchmark figures are listed in section \ref{sec} .}
 \label{fig:benchmark}
\end{figure}
\par

\begin{figure}
 \centering
 \includegraphics[width=\linewidth]{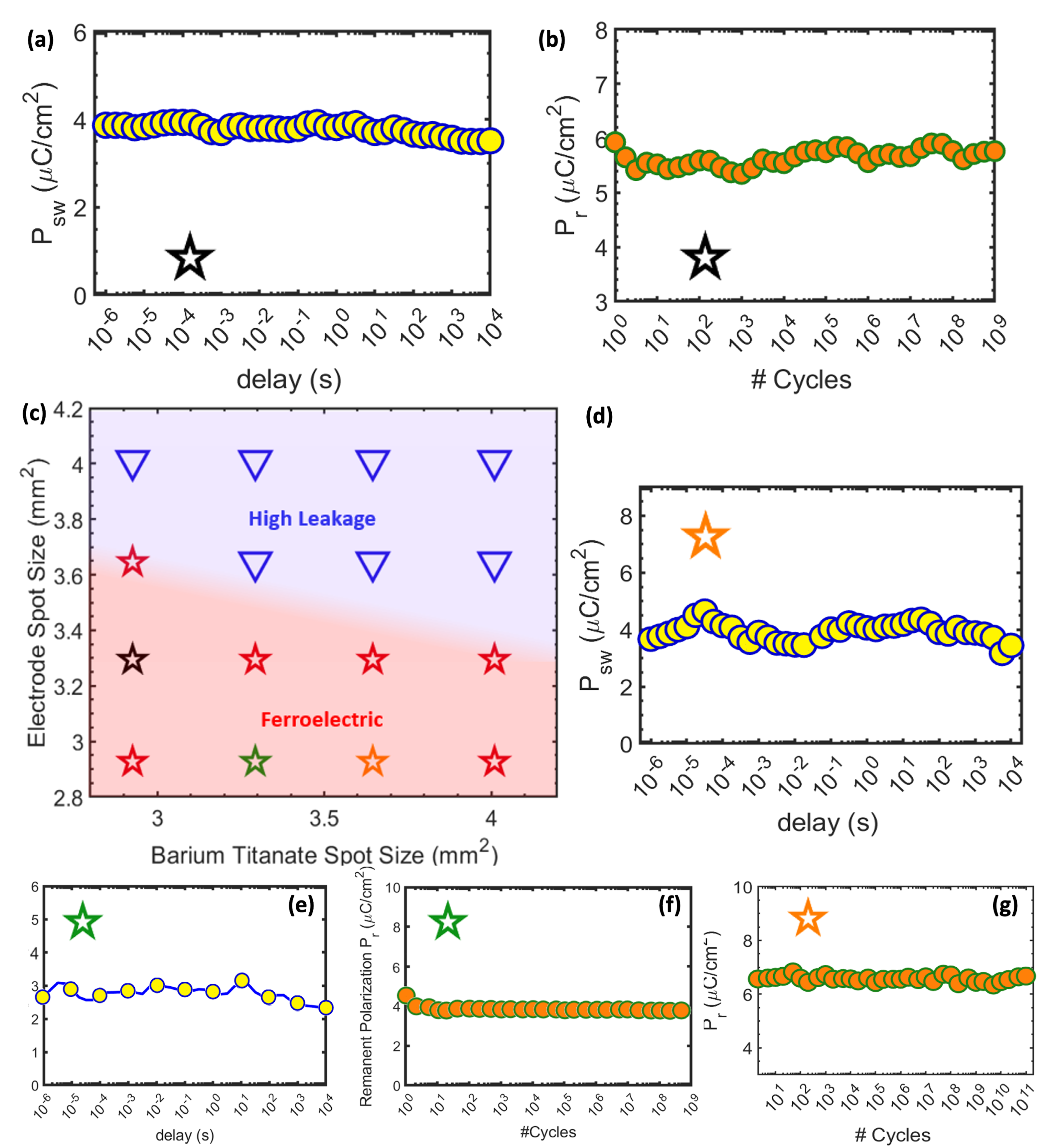}
 \captionof{figure}{Endurance and Retention characteristics for multiple 20nm SRO/ 20nm BTO/ 20nm SRO/GSO heterostructures. The location in the phasespace is indicated by a colour. All retention characteristics were measured using a series of pulses as shown in the inset of figure 4(a). Heterostructures where in endurance characteristics were measured upto 10$^9$ cycles, used a continuous square pulse of 0.5V amplitude used a 10$\mu$s pulse width. The heterostructure where in 10$^{11}$ cycles were measured was subject to a continuous square pulse of 0.7V amplitude used a 100ns pulse width.}
 \label{fig:RetnandEndr_multfilms}
\end{figure}
\par

\vspace{6pt}

Having established the necessity and importance of long-term retention in ultrathin ferroelectrics that simultaneously switch at a low voltage, it is pertinent to look at the literature of their retention characteristics.  Figure \ref{fig:benchmark}(a)  compares retention times as a function of coercive voltages among oxide ferroelectrics from other studies, with our work and it can be seen that all films that show stable retention of at least $10^4$ seconds have a coercive voltage of 1V or higher. Intuitively, a larger coercive voltage implies a higher barrier height for the spontaneous polarization against relaxation. In the past, a sufficiently thick perovskite oxide film was necessary to show long retention times because of their lower leakage currents and larger coercive voltages. If we want to exploit the low coercive fields of perovskites and still achieve long-term retention, we will want to access regimes of long-term-stable polarization states that switch under very low voltages. The films produced in this work occupy this elusive section of the phase space.

\par

In addition to non-volatility, an associated parameter is the endurance, which is the number of cycles to breakdown of the ferroelectric film. Figure \ref{fig:benchmark}(b) compares the endurance properties of our film with those of other fluorite and perovskite oxide ferroelectrics. There are very few reports of fluorite oxides showing fatigue free nature above $10^{10}$ cycles because of their close to breakdown field cycling. Whereas perovskite oxides show no fatigue even up to $10^{12}$ cycles albeit at much greater thicknesses. In the earlier days, the fatigue in perovskite oxides were primarily due to a dead layer formation and the accumulation of oxygen vacancies at the electrode-ferroelectric interface, both of which were solved by adopting conducting oxide electrodes. Nevertheless, it is intriguing that there are hardly any reports of endurance measurements in sub-50 nm films. It is unclear if thinner films should intrinsically have worse endurance characteristics or if there have not been many endurance measurements on ultra-thin perovskite films considering the difficulty in stabilizing ferroelectricity in these films. One possible reason for thinner films having worse endurance characteristics could be that as the ferroelectric becomes thinner, the interface to bulk ratio increases and the probability of domain-pinning due to an interface defect/misfit dislocation also increases. Nevertheless, the films from this study simultaneously possess unprecedented endurance and retention characteristics. Figure \ref{fig:RetnandEndr_multfilms} also lists the retention and endurance characteristics measured in some of the different 20nm SRO/ 20nm BTO/ 20nm SRO/GSO heterostructures from this study.

\section{References for Retention and Endurance benchmarks}\label{sec}

Retention times from figure \ref{fig:benchmark}(a): Orange triangles-fluorite oxides\cite{lyu2020high,zeng20192,sunbul2023impact,peng2021hfo,liang2022zro} and purple diamonds - perovskite oxide ferroelectrics \cite{dat1994polycrystalline,dhote1996direct,yang1997low,ramesh1994oriented,ramesh1993ferroelectric,ramesh1992fatigue,wang2004epitaxial,aggarwal1999switching,nagaraj2001influence,ganpule1999scaling,jiang2022enabling,tan2023improved,bakaul2017high,yoo2005highly,jo2005thickness}.

Endurance data from figure \ref{fig:benchmark}(b): orange circles-fluorite oxides\cite{cao2019improvement,migita2018polarization,schroeder2013hafnium,chernikova2018improved,lyu2020high,gaddam2020insertion,lo2021high,kozodaev2019mitigating,toprasertpong2022low,li2020involvement,liu2018endurance,sunbul2023impact,peng2021hfo,liang2022zro} and purple squares - perovskite oxide ferroelectrics\cite{singh2006improved,dat1994polycrystalline,shimizu1997effects,dhote1996direct,yang1997low,ramesh1994oriented,ramesh1993ferroelectric,ramesh1992fatigue,maki2003controlling,rodriguez2004reliability,guo2013non,jiang2022enabling,muller2021training,bakaul2017high,watanabe2001film,yoo2005highly}

\newpage

\bibliographystyle{ieeetr}
\renewcommand{\bibfont}{\small}
\bibliography{thesis}

\begin{thebibliography}{10}

\bibitem{dawber2005physics}
M.~Dawber, K.~Rabe, and J.~Scott, ``{Physics of thin-film ferroelectric oxides},'' {\em Reviews of modern physics}, vol.~77, no.~4, p.~1083, 2005.

\bibitem{stengel2006origin}
M.~Stengel and N.~A. Spaldin, ``{Origin of the dielectric dead layer in nanoscale capacitors},'' {\em Nature}, vol.~443, no.~7112, pp.~679--682, 2006.

\bibitem{fernandez2022thin}
A.~Fernandez, M.~Acharya, H.-G. Lee, J.~Schimpf, Y.~Jiang, D.~Lou, Z.~Tian, and L.~W. Martin, ``{Thin-film ferroelectrics},'' {\em Advanced materials}, vol.~34, no.~30, p.~2108841, 2022.

\bibitem{sai2005ferroelectricity}
N.~Sai, A.~M. Kolpak, and A.~M. Rappe, ``{Ferroelectricity in ultrathin perovskite films},'' {\em Physical Review B}, vol.~72, no.~2, p.~020101, 2005.

\bibitem{junquera2003critical}
J.~Junquera and P.~Ghosez, ``{Critical thickness for ferroelectricity in perovskite ultrathin films},'' {\em Nature}, vol.~422, no.~6931, pp.~506--509, 2003.

\bibitem{kim2005polarization}
D.~Kim, J.~Jo, Y.~Kim, Y.~Chang, J.~Lee, J.-G. Yoon, T.~Song, and T.~Noh, ``{Polarization relaxation induced by a depolarization field in ultrathin ferroelectric BaTiO$_3$ capacitors},'' {\em Physical review letters}, vol.~95, no.~23, p.~237602, 2005.

\bibitem{ma2002nonvolatile}
T.~Ma and J.-P. Han, ``{Why is nonvolatile ferroelectric memory field-effect transistor still elusive?},'' {\em IEEE Electron Device Letters}, vol.~23, no.~7, pp.~386--388, 2002.

\bibitem{kim2023wurtzite}
K.-H. Kim, I.~Karpov, R.~H. Olsson~III, and D.~Jariwala, ``{Wurtzite and fluorite ferroelectric materials for electronic memory},'' {\em Nature Nanotechnology}, vol.~18, no.~5, pp.~422--441, 2023.

\bibitem{mikolajick2021next}
T.~Mikolajick, S.~Slesazeck, H.~Mulaosmanovic, M.~Park, S.~Fichtner, P.~Lomenzo, M.~Hoffmann, and U.~Schroeder, ``{Next generation ferroelectric materials for semiconductor process integration and their applications},'' {\em Journal of Applied Physics}, vol.~129, no.~10, 2021.

\bibitem{hoffmann2021progress}
M.~Hoffmann, S.~Slesazeck, and T.~Mikolajick, ``{Progress and future prospects of negative capacitance electronics: A materials perspective},'' {\em APL Materials}, vol.~9, no.~2, 2021.

\bibitem{jiang2022enabling}
Y.~Jiang, E.~Parsonnet, A.~Qualls, W.~Zhao, S.~Susarla, D.~Pesquera, A.~Dasgupta, M.~Acharya, H.~Zhang, T.~Gosavi, {\em et~al.}, ``{Enabling ultra-low-voltage switching in BaTiO$_3$},'' {\em Nature materials}, vol.~21, no.~7, pp.~779--785, 2022.

\bibitem{manipatruni2018beyond}
S.~Manipatruni, D.~E. Nikonov, and I.~A. Young, ``{Beyond CMOS computing with spin and polarization},'' {\em Nature Physics}, vol.~14, no.~4, pp.~338--343, 2018.

\bibitem{gao2017possible}
P.~Gao, Z.~Zhang, M.~Li, R.~Ishikawa, B.~Feng, H.-J. Liu, Y.-L. Huang, N.~Shibata, X.~Ma, S.~Chen, {\em et~al.}, ``{Possible absence of critical thickness and size effect in ultrathin perovskite ferroelectric films},'' {\em Nature communications}, vol.~8, no.~1, p.~15549, 2017.

\bibitem{lee2019first}
S.~R. Lee, L.~Baasandorj, J.~W. Chang, I.~W. Hwang, J.~R. Kim, J.-G. Kim, K.-T. Ko, S.~B. Shim, M.~W. Choi, M.~You, {\em et~al.}, ``{First observation of ferroelectricity in $ \sim 1$ nm ultrathin semiconducting BaTiO$_3$ films},'' {\em Nano letters}, vol.~19, no.~4, pp.~2243--2250, 2019.

\bibitem{nagarajan2006scaling}
V.~Nagarajan, J.~Junquera, J.~He, C.~Jia, R.~Waser, K.~Lee, Y.~Kim, S.~Baik, T.~Zhao, R.~Ramesh, {\em et~al.}, ``{Scaling of structure and electrical properties in ultrathin epitaxial ferroelectric heterostructures},'' {\em Journal of applied physics}, vol.~100, no.~5, 2006.

\bibitem{shaw2000properties}
T.~Shaw, S.~Trolier-McKinstry, and P.~McIntyre, ``{The properties of ferroelectric films at small dimensions},'' {\em Annual Review of Materials Science}, vol.~30, no.~1, pp.~263--298, 2000.

\bibitem{jo2005thickness}
J.~Jo, Y.~Kim, D.~Kim, J.~Kim, Y.~Chang, J.~Kong, Y.~Park, T.~Song, J.-G. Yoon, J.~Jung, {\em et~al.}, ``{Thickness-dependent ferroelectric properties in fully-strained SrRuO$_3$/BaTiO$_3$/SrRuO$_3$ ultra-thin capacitors},'' {\em Thin Solid Films}, vol.~486, no.~1-2, pp.~149--152, 2005.

\bibitem{seol2017non}
D.~Seol, B.~Kim, and Y.~Kim, ``{Non-piezoelectric effects in piezoresponse force microscopy},'' {\em Current Applied Physics}, vol.~17, no.~5, pp.~661--674, 2017.

\bibitem{vasudevan2017ferroelectric}
R.~K. Vasudevan, N.~Balke, P.~Maksymovych, S.~Jesse, and S.~V. Kalinin, ``{Ferroelectric or non-ferroelectric: Why so many materials exhibit “ferroelectricity” on the nanoscale},'' {\em Applied Physics Reviews}, vol.~4, no.~2, 2017.

\bibitem{gong2016fe}
N.~Gong and T.-P. Ma, ``{Why is FE--HfO$_2$ more suitable than PZT or SBT for scaled nonvolatile 1-T memory cell? A retention perspective},'' {\em IEEE Electron Device Letters}, vol.~37, no.~9, pp.~1123--1126, 2016.

\bibitem{black1999electric}
C.~T. Black and J.~J. Welser, ``{Electric-field penetration into metals: consequences for high-dielectric-constant capacitors},'' {\em IEEE Transactions on Electron Devices}, vol.~46, no.~4, pp.~776--780, 1999.

\bibitem{martin2024lifting}
L.~W. Martin, J.-P. Maria, and D.~G. Schlom, ``{Lifting the fog in ferroelectric thin-film synthesis},'' {\em Nature Materials}, pp.~1--2, 2024.

\bibitem{xu2013impact}
C.~Xu, S.~Wicklein, A.~Sambri, S.~Amoruso, M.~Moors, and R.~Dittmann, ``{Impact of the interplay between nonstoichiometry and kinetic energy of the plume species on the growth mode of SrTiO$_3$ thin films},'' {\em Journal of Physics D: Applied Physics}, vol.~47, no.~3, p.~034009, 2013.

\bibitem{scullin2010pulsed}
M.~L. Scullin, J.~Ravichandran, C.~Yu, M.~Huijben, J.~Seidel, A.~Majumdar, and R.~Ramesh, ``{Pulsed laser deposition-induced reduction of SrTiO$_3$ crystals},'' {\em Acta materialia}, vol.~58, no.~2, pp.~457--463, 2010.

\bibitem{thompson2016enhanced}
J.~Thompson, J.~Nichols, S.~Lee, S.~Ryee, J.~H. Gruenewald, J.~G. Connell, M.~Souri, J.~Johnson, J.~Hwang, M.~J. Han, {\em et~al.}, ``{Enhanced metallic properties of SrRuO$_3$ thin films via kinetically controlled pulsed laser epitaxy},'' {\em Applied Physics Letters}, vol.~109, no.~16, 2016.

\bibitem{saremi2016enhanced}
S.~Saremi, R.~Xu, L.~R. Dedon, J.~A. Mundy, S.-L. Hsu, Z.~Chen, A.~R. Damodaran, S.~P. Chapman, J.~T. Evans, and L.~W. Martin, ``{Enhanced electrical resistivity and properties via ion bombardment of ferroelectric thin films},'' {\em Advanced materials}, vol.~28, no.~48, pp.~10750--10756, 2016.

\bibitem{biegalski2005thermal}
M.~Biegalski, J.~Haeni, S.~Trolier-McKinstry, D.~Schlom, C.~Brandle, and A.~V. Graitis, ``{Thermal expansion of the new perovskite substrates DyScO$_3$ and GdScO$_3$},'' {\em Journal of Materials Research}, vol.~20, no.~4, pp.~952--958, 2005.

\bibitem{kan2011controlled}
D.~Kan and Y.~Shimakawa, ``{Controlled cation stoichiometry in pulsed laser deposition-grown BaTiO$_3$ epitaxial thin films with laser fluence},'' {\em Applied Physics Letters}, vol.~99, no.~8, 2011.

\bibitem{chen2013strong}
A.~Chen, F.~Khatkhatay, W.~Zhang, C.~Jacob, L.~Jiao, and H.~Wang, ``{Strong oxygen pressure dependence of ferroelectricity in BaTiO$_3$/SrRuO$_3$/SrTiO$_3$ epitaxial heterostructures},'' {\em Journal of Applied Physics}, vol.~114, no.~12, 2013.

\bibitem{harris2020geometrical}
S.~B. Harris, K.~L. Kopekcy, C.~W. Cotton, and R.~P. Camata, ``{Geometrical and energy scaling in the pulsed laser deposition plasma during epitaxial growth of FeSe thin films},'' {\em arXiv preprint arXiv:2002.09701}, 2020.

\bibitem{lee2016growth}
H.~N. Lee, S.~S. Ambrose~Seo, W.~S. Choi, and C.~M. Rouleau, ``{Growth control of oxygen stoichiometry in homoepitaxial SrTiO$_3$ films by pulsed laser epitaxy in high vacuum},'' {\em Scientific reports}, vol.~6, no.~1, p.~19941, 2016.

\bibitem{wan2019nonvolatile}
S.~Wan, Y.~Li, W.~Li, X.~Mao, C.~Wang, C.~Chen, J.~Dong, A.~Nie, J.~Xiang, Z.~Liu, {\em et~al.}, ``{Nonvolatile ferroelectric memory effect in ultrathin $\alpha$-In$_2$Se$_3$},'' {\em Advanced Functional Materials}, vol.~29, no.~20, p.~1808606, 2019.

\bibitem{mathews1997ferroelectric}
S.~Mathews, R.~Ramesh, T.~Venkatesan, and J.~Benedetto, ``Ferroelectric field effect transistor based on epitaxial perovskite heterostructures,'' {\em Science}, vol.~276, no.~5310, pp.~238--240, 1997.

\bibitem{watanabe1995epitaxial}
Y.~Watanabe, ``{Epitaxial all-perovskite ferroelectric field effect transistor with a memory retention},'' {\em Applied physics letters}, vol.~66, no.~14, pp.~1770--1772, 1995.

\bibitem{aizawa2004impact}
K.~Aizawa, B.-E. Park, Y.~Kawashima, K.~Takahashi, and H.~Ishiwara, ``{Impact of HfO$_2$ buffer layers on data retention characteristics of ferroelectric-gate field-effect transistors},'' {\em Applied physics letters}, vol.~85, no.~15, pp.~3199--3201, 2004.

\bibitem{hoffman2011device}
J.~Hoffman, X.~Hong, and C.~Ahn, ``{Device performance of ferroelectric/correlated oxide heterostructures for non-volatile memory applications},'' {\em Nanotechnology}, vol.~22, no.~25, p.~254014, 2011.

\bibitem{kim2005critical}
Y.~Kim, D.~Kim, J.~Kim, Y.~Chang, T.~Noh, J.~Kong, K.~Char, Y.~Park, S.~Bu, J.-G. Yoon, {\em et~al.}, ``{Critical thickness of ultrathin ferroelectric BaTiO$_3$ films},'' {\em Applied Physics Letters}, vol.~86, no.~10, p.~102907, 2005.

\bibitem{jo2006polarization}
J.~Jo, D.~Kim, Y.~Kim, S.-B. Choe, T.~Song, J.-G. Yoon, and T.~Noh, ``{Polarization switching dynamics governed by the thermodynamic nucleation process in ultrathin ferroelectric films},'' {\em Physical review letters}, vol.~97, no.~24, p.~247602, 2006.

\bibitem{takahashi2005thirty}
K.~Takahashi, K.~Aizawa, B.-E. Park, and H.~Ishiwara, ``{Thirty-day-long data retention in ferroelectric-gate field-effect transistors with HfO$_2$ buffer layers},'' {\em Japanese journal of applied physics}, vol.~44, no.~8R, p.~6218, 2005.

\bibitem{weber2013variable}
D.~Weber, R.~V{\H{o}}f{\'e}ly, Y.~Chen, Y.~Mourzina, and U.~Poppe, ``{Variable resistor made by repeated steps of epitaxial deposition and lithographic structuring of oxide layers by using wet chemical etchants},'' {\em Thin Solid Films}, vol.~533, pp.~43--47, 2013.

\bibitem{mehta1973depolarization}
R.~Mehta, B.~Silverman, and J.~Jacobs, ``Depolarization fields in thin ferroelectric films,'' {\em Journal of Applied Physics}, vol.~44, no.~8, pp.~3379--3385, 1973.

\bibitem{lyu2020high}
J.~Lyu, T.~Song, I.~Fina, and F.~S{\'a}nchez, ``{High polarization, endurance and retention in sub-5 nm Hf$_{0.5}$Zr$_{0.5}$O$_2$ films},'' {\em Nanoscale}, vol.~12, no.~20, pp.~11280--11287, 2020.

\bibitem{zeng20192}
B.~Zeng, M.~Liao, Q.~Peng, W.~Xiao, J.~Liao, S.~Zheng, and Y.~Zhou, ``{2-bit/cell operation of Hf$_{0.5}$Zr$_{0.5}$O$_2$ based FeFET memory devices for NAND applications},'' {\em IEEE Journal of the Electron Devices Society}, vol.~7, pp.~551--556, 2019.

\bibitem{sunbul2023impact}
A.~S{\"u}nb{\"u}l, D.~Lehninger, R.~Hoffmann, R.~Olivo, A.~Prabhu, F.~Sch{\"o}ne, K.~K{\"u}hnel, M.~D{\"o}llgast, N.~Haufe, L.~Roy, {\em et~al.}, ``{Impact of Ferroelectric Layer Thickness on Reliability of Back-End-of-Line-Compatible Hafnium Zirconium Oxide Films},'' {\em Advanced Engineering Materials}, vol.~25, no.~4, p.~2201124, 2023.

\bibitem{peng2021hfo}
Y.~Peng, W.~Xiao, Y.~Liu, C.~Jin, X.~Deng, Y.~Zhang, F.~Liu, Y.~Zheng, Y.~Cheng, B.~Chen, {\em et~al.}, ``{HfO$_2$-ZrO$_2$ superlattice ferroelectric capacitor with improved endurance performance and higher fatigue recovery capability},'' {\em IEEE Electron Device Letters}, vol.~43, no.~2, pp.~216--219, 2021.

\bibitem{liang2022zro}
Y.-K. Liang, W.-L. Li, Y.-J. Wang, L.-C. Peng, C.-C. Lu, H.-Y. Huang, S.~H. Yeong, Y.-M. Lin, Y.-H. Chu, E.-Y. Chang, {\em et~al.}, ``{ZrO$_2$-HfO$_2$ Superlattice Ferroelectric Capacitors with Optimized Annealing to Achieve Extremely High Polarization Stability},'' {\em IEEE Electron Device Letters}, vol.~43, no.~9, pp.~1451--1454, 2022.

\bibitem{dat1994polycrystalline}
R.~Dat, D.~Lichtenwalner, O.~Auciello, and A.~Kingon, ``{Polycrystalline La$_{0. 5}$Sr$_{0. 5}$CoO$_3$/PbZr$_{0. 53}$Ti$_{0. 47}$O$_3$/La$_{0. 5}$Sr$_{0. 5}$CoO$_3$ ferroelectric capacitors on platinized silicon with no polarization fatigue},'' {\em Applied physics letters}, vol.~64, no.~20, pp.~2673--2675, 1994.

\bibitem{dhote1996direct}
A.~Dhote, S.~Madhukar, W.~Wei, T.~Venkatesan, R.~Ramesh, and C.~Cotell, ``{Direct integration of ferroelectric La--Sr--Co--O/Pb--Nb--Zr--Ti--O/La--Sr--Co--O capacitors on silicon with conducting barrier layers},'' {\em Applied physics letters}, vol.~68, no.~10, pp.~1350--1352, 1996.

\bibitem{yang1997low}
B.~Yang, T.~Song, S.~Aggarwal, and R.~Ramesh, ``{Low voltage performance of Pb(Zr, Ti)O$_3$ capacitors through donor doping},'' {\em Applied physics letters}, vol.~71, no.~24, pp.~3578--3580, 1997.

\bibitem{ramesh1994oriented}
R.~Ramesh, J.~Lee, T.~Sands, V.~Keramidas, and O.~Auciello, ``{Oriented ferroelectric La-Sr-Co-O/Pb-La-Zr-Ti-O/La-Sr-Co-O heterostructures on [001] Pt/SiO$_2$ Si substrates using a bismuth titanate template layer},'' {\em Applied physics letters}, vol.~64, no.~19, pp.~2511--2513, 1994.

\bibitem{ramesh1993ferroelectric}
R.~Ramesh, H.~Gilchrist, T.~Sands, V.~Keramidas, R.~Haakenaasen, and D.~Fork, ``{Ferroelectric La-Sr-Co-O/Pb-Zr-Ti-O/La-Sr-Co-O heterostructures on silicon via template growth},'' {\em Applied Physics Letters}, vol.~63, no.~26, pp.~3592--3594, 1993.

\bibitem{ramesh1992fatigue}
R.~Ramesh, W.~Chan, B.~Wilkens, H.~Gilchrist, T.~Sands, J.~Tarascon, V.~Keramidas, D.~Fork, J.~Lee, and A.~Safari, ``{Fatigue and retention in ferroelectric Y-Ba-Cu-O/Pb-Zr-Ti-O/Y-Ba-Cu-O heterostructures},'' {\em Applied Physics Letters}, vol.~61, no.~13, pp.~1537--1539, 1992.

\bibitem{wang2004epitaxial}
J.~Wang, H.~Zheng, Z.~Ma, S.~Prasertchoung, M.~Wuttig, R.~Droopad, J.~Yu, K.~Eisenbeiser, and R.~Ramesh, ``{Epitaxial BiFeO$_3$ thin films on Si},'' {\em Applied Physics Letters}, vol.~85, no.~13, pp.~2574--2576, 2004.

\bibitem{aggarwal1999switching}
S.~Aggarwal, I.~Jenkins, B.~Nagaraj, C.~Kerr, C.~Canedy, R.~Ramesh, G.~Velasquez, L.~Boyer, and J.~Evans~Jr, ``{Switching properties of Pb (Nb, Zr, Ti)O$_3$ capacitors using SrRuO$_3$ electrodes},'' {\em Applied physics letters}, vol.~75, no.~12, pp.~1787--1789, 1999.

\bibitem{nagaraj2001influence}
B.~Nagaraj, S.~Aggarwal, and R.~Ramesh, ``{Influence of contact electrodes on leakage characteristics in ferroelectric thin films},'' {\em Journal of applied physics}, vol.~90, no.~1, pp.~375--382, 2001.

\bibitem{ganpule1999scaling}
C.~Ganpule, A.~Stanishevsky, Q.~Su, S.~Aggarwal, J.~Melngailis, E.~Williams, and R.~Ramesh, ``{Scaling of ferroelectric properties in thin films},'' {\em Applied physics letters}, vol.~75, no.~3, pp.~409--411, 1999.

\bibitem{tan2023improved}
X.~Tan, X.~Sun, J.~Jiang, and D.~Chen, ``{Improved polarization retention in epitaxial BiFeO$_3$ thin films induced by strain relaxation},'' {\em Applied Surface Science}, p.~157703, 2023.

\bibitem{bakaul2017high}
S.~R. Bakaul, C.~R. Serrao, O.~Lee, Z.~Lu, A.~Yadav, C.~Carraro, R.~Maboudian, R.~Ramesh, and S.~Salahuddin, ``{High Speed Epitaxial Perovskite Memory on Flexible Substrates.},'' {\em Advanced Materials (Deerfield Beach, Fla.)}, vol.~29, no.~11, 2017.

\bibitem{yoo2005highly}
D.~Yoo, B.~Bae, J.~Lim, D.~Im, S.~Park, S.~Kim, U.-I. Chung, J.~Moon, and B.~Ryu, ``{Highly reliable 50nm-thick PZT capacitor and low voltage FRAM device using Ir/SrRuO$_3$/MOCVD PZT capacitor technology},'' in {\em Digest of Technical Papers. 2005 Symposium on VLSI Technology, 2005.}, pp.~100--101, IEEE, 2005.

\bibitem{cao2019improvement}
R.~Cao, B.~Song, D.~Shang, Y.~Yang, Q.~Luo, S.~Wu, Y.~Li, Y.~Wang, H.~Lv, Q.~Liu, {\em et~al.}, ``{Improvement of endurance in HZO-based ferroelectric capacitor using Ru electrode},'' {\em IEEE Electron Device Letters}, vol.~40, no.~11, pp.~1744--1747, 2019.

\bibitem{migita2018polarization}
S.~Migita, H.~Ota, H.~Yamada, K.~Shibuya, A.~Sawa, and A.~Toriumi, ``{Polarization switching behavior of Hf--Zr--O ferroelectric ultrathin films studied through coercive field characteristics},'' {\em Japanese Journal of Applied Physics}, vol.~57, no.~4S, p.~04FB01, 2018.

\bibitem{schroeder2013hafnium}
U.~Schroeder, S.~Mueller, J.~Mueller, E.~Yurchuk, D.~Martin, C.~Adelmann, T.~Schloesser, R.~van Bentum, and T.~Mikolajick, ``{Hafnium oxide based CMOS compatible ferroelectric materials},'' {\em ECS Journal of Solid State Science and Technology}, vol.~2, no.~4, p.~N69, 2013.

\bibitem{chernikova2018improved}
A.~G. Chernikova, M.~G. Kozodaev, D.~V. Negrov, E.~V. Korostylev, M.~H. Park, U.~Schroeder, C.~S. Hwang, and A.~M. Markeev, ``{Improved ferroelectric switching endurance of La-doped Hf$_{0.5}$Zr$_{0.5}$O$_2$ thin films},'' {\em ACS applied materials \& interfaces}, vol.~10, no.~3, pp.~2701--2708, 2018.

\bibitem{gaddam2020insertion}
V.~Gaddam, D.~Das, and S.~Jeon, ``{Insertion of HfO$_2$ seed/dielectric layer to the ferroelectric HZO films for heightened remanent polarization in MFM capacitors},'' {\em IEEE Transactions on Electron Devices}, vol.~67, no.~2, pp.~745--750, 2020.

\bibitem{lo2021high}
C.~Lo, C.-K. Chen, C.-F. Chang, F.-S. Zhang, Z.-H. Lu, and T.-S. Chao, ``{High endurance and low fatigue effect of bilayer stacked antiferroelectric/ferroelectric Hf$_{x}$Zr$_{1-x}$O$_2$},'' {\em IEEE Electron Device Letters}, vol.~43, no.~2, pp.~224--227, 2021.

\bibitem{kozodaev2019mitigating}
M.~G. Kozodaev, A.~G. Chernikova, E.~V. Korostylev, M.~H. Park, R.~R. Khakimov, C.~S. Hwang, and A.~M. Markeev, ``{Mitigating wakeup effect and improving endurance of ferroelectric HfO$_2$-ZrO$_2$ thin films by careful La-doping},'' {\em Journal of Applied Physics}, vol.~125, no.~3, 2019.

\bibitem{toprasertpong2022low}
K.~Toprasertpong, K.~Tahara, Y.~Hikosaka, K.~Nakamura, H.~Saito, M.~Takenaka, and S.~Takagi, ``{Low operating voltage, improved breakdown tolerance, and high endurance in Hf$_{0.5}$Zr$_{0.5}$O$_2$ ferroelectric capacitors achieved by thickness scaling down to 4 nm for embedded ferroelectric memory},'' {\em ACS Applied Materials \& Interfaces}, vol.~14, no.~45, pp.~51137--51148, 2022.

\bibitem{li2020involvement}
S.~Li, D.~Zhou, Z.~Shi, M.~Hoffmann, T.~Mikolajick, and U.~Schroeder, ``{Involvement of unsaturated switching in the endurance cycling of Si-doped HfO$_2$ ferroelectric thin films},'' {\em Advanced Electronic Materials}, vol.~6, no.~8, p.~2000264, 2020.

\bibitem{liu2018endurance}
X.~Liu, D.~Zhou, Y.~Guan, S.~Li, F.~Cao, and X.~Dong, ``{Endurance properties of silicon-doped hafnium oxide ferroelectric and antiferroelectric-like thin films: A comparative study and prediction},'' {\em Acta Materialia}, vol.~154, pp.~190--198, 2018.

\bibitem{singh2006improved}
S.~Singh and H.~Ishiwara, ``{Improved fatigue endurance in Mn-doped Bi$_{3.25}$La$_{0. 75}$Ti$_3$O$_{12}$ thin films},'' {\em Solid state communications}, vol.~140, no.~9-10, pp.~430--434, 2006.

\bibitem{shimizu1997effects}
M.~Shimizu, H.~Fujisawa, S.~Hyodo, S.~Nakashima, H.~Niu, H.~Okino, and T.~Shiosaki, ``{Effects of Sputtered Ir and IrO$_2$ Electrodes on the Properties of PZT Thin Films Deposited By MOCVD},'' {\em MRS Online Proceedings Library (OPL)}, vol.~493, 1997.

\bibitem{maki2003controlling}
K.~Maki, B.~Liu, H.~Vu, V.~Nagarajan, R.~Ramesh, Y.~Fujimori, T.~Nakamura, and H.~Takasu, ``{Controlling crystallization of Pb(Zr,Ti)O$_3$ thin films on IrO$_2$ electrodes at low temperature through interface engineering},'' {\em Applied physics letters}, vol.~82, no.~8, pp.~1263--1265, 2003.

\bibitem{rodriguez2004reliability}
J.~A. Rodriguez, K.~Remack, K.~Boku, K.~Udayakumar, S.~Aggarwal, S.~R. Summerfelt, F.~G. Celii, S.~Martin, L.~Hall, K.~Taylor, {\em et~al.}, ``{Reliability properties of low-voltage ferroelectric capacitors and memory arrays},'' {\em IEEE Transactions on Device and Materials Reliability}, vol.~4, no.~3, pp.~436--449, 2004.

\bibitem{guo2013non}
R.~Guo, L.~You, Y.~Zhou, Z.~Shiuh~Lim, X.~Zou, L.~Chen, R.~Ramesh, and J.~Wang, ``{Non-volatile memory based on the ferroelectric photovoltaic effect},'' {\em Nature communications}, vol.~4, no.~1, p.~1990, 2013.

\bibitem{muller2021training}
M.~M{\"u}ller, Y.-L. Huang, S.~V{\'e}lez, R.~Ramesh, M.~Fiebig, and M.~Trassin, ``{Training the Polarization in Integrated La$_{0. 15}$Bi$_{0. 85}$FeO$_3$-Based Devices},'' {\em Advanced Materials}, vol.~33, no.~52, p.~2104688, 2021.

\bibitem{watanabe2001film}
T.~Watanabe, A.~Saiki, K.~Saito, and H.~Funakubo, ``{Film thickness dependence of ferroelectric properties of c-axis-oriented epitaxial Bi$_4$Ti$_3$O$_{12}$ thin films prepared by metalorganic chemical vapor deposition},'' {\em Journal of Applied Physics}, vol.~89, no.~7, pp.~3934--3938, 2001.

\end{thebibliography}

\end{document}